\documentclass[tighten, notrackchanges, twocolumn]{aastex631}
\usepackage{amsmath}
\usepackage{amssymb}
\usepackage{amssymb}
\usepackage{mathtools}
\usepackage{mathrsfs}
\usepackage{enumitem}
\usepackage[T1]{fontenc}
\usepackage{morefloats}
\usepackage{longtable}
\usepackage{afterpage}
\usepackage{acronym}   
\hypersetup{linkcolor=red,citecolor=blue,urlcolor=magenta}
\usepackage{color,xcolor, colortbl}
\usepackage{array, booktabs, boldline}
\definecolor{LightGreen}{rgb}{0.714,0.965,0.722}
\usepackage{soul}

\usepackage[makeroom]{cancel}

\graphicspath{{figs/}}

\usepackage{xspace}

\newcommand*{\XPSI}{X-PSI\xspace}
\newcommand*{\NICER}{NICER\xspace}
\newcommand*{\xmm}{XMM-Newton\xspace}
\newcommand*{\MultiNest}{\textsc{MultiNest}\xspace}
\newcommand*{\PyMultiNest}{\textsc{PyMultiNest}\xspace}

\newcommand{\joo}{PSR~J0030$+$0451\xspace}
\newcommand{\joh}{PSR~J0740$+$6620\xspace}

\acrodef{X-PSI}{X-ray Pulse Simulation and Inference}
\acrodef{MSP}{millisecond pulsar}
\acrodef{PPM}{Pulse Profile Modeling}
\acrodef{EoS}{Equation of State}
\acrodef{ISS}{International Space Station}
\acrodef{NS}{neutron star}
\acrodef{KDE}{Kernel Density Estimation}
\acrodefplural{NS}[NSs]{neutron stars}
\acrodef{NICER}{Neutron Star Interior Composition Explorer}
\acrodef{GTI}{Good Time Interval}
\acrodefplural{GTI}[GTIs]{Good Time Intervals}
\acrodefplural{KDE}[KDEs]{Kernel Density Estimations}

\newcommand{\be}{\begin{equation}}
\newcommand{\ee}{\end{equation}}

\defcitealias{Riley2019}{R19}
\defcitealias{Bogdanov2019}{B19}
\defcitealias{Vinciguerra2022a}{V22a}

\received{}
\revised{}
\accepted{}
\published{}

\submitjournal{nowhere}

\shorttitle{Simulations for PSR~J0030+0451}
\shortauthors{Vinciguerra~et~al.}

\begin{document}

\title{X-PSI Parameter Recovery for Temperature Map Configurations Inspired by PSR J0030+0451 \footnote{Submitted on -, -, 2023}}

\correspondingauthor{S.~Vinciguerra}
\email{s.vinciguerra@uva.nl}

\author[0000-0003-3068-6974]{Serena~Vinciguerra}
\affil{Anton Pannekoek Institute for Astronomy, University of Amsterdam, Science Park 904, 1090GE Amsterdam, the Netherlands}
https://www.overleaf.com/project/63331d06041e4510858c5dda
\author[0000-0001-6356-125X ]{Tuomo~Salmi}
\affil{Anton Pannekoek Institute for Astronomy, University of Amsterdam, Science Park 904, 1090GE Amsterdam, the Netherlands}

\author[0000-0002-1009-2354]{Anna~L.~Watts}
\affil{Anton Pannekoek Institute for Astronomy, University of Amsterdam, Science Park 904, 1090GE Amsterdam, the Netherlands}

\author[0000-0002-2651-5286]{Devarshi~Choudhury}
\affil{Anton Pannekoek Institute for Astronomy, University of Amsterdam, Science Park 904, 1090GE Amsterdam, the Netherlands}

\author[0000-0002-0428-8430]{Yves~Kini}
\affil{Anton Pannekoek Institute for Astronomy, University of Amsterdam, Science Park 904, 1090GE Amsterdam, the Netherlands}

\author[0000-0001-9313-0493]{Thomas~E.~Riley}
\affil{Anton Pannekoek Institute for Astronomy, University of Amsterdam, Science Park 904, 1090GE Amsterdam, the Netherlands}

\begin{abstract}
In the last few years, the \NICER collaboration has 
provided 
mass and radius inferences, via pulse profile modeling, for two pulsars: \joo and \joh. 
Given the importance of these results for constraining the equation of state of dense nuclear matter, 
it is crucial to validate them and test their robustness. 
We therefore explore the reliability of these results and their sensitivity to analysis settings and random processes, including noise, focusing on the specific case of \joo. 
We use \XPSI, one of the two main analysis pipelines currently employed by the \NICER collaboration for mass and radius inferences. 
With synthetic data that mimic the \joo \NICER data set, we evaluate the recovery performances of \XPSI under conditions never tested before, including complex modeling of the thermally emitting neutron star surface. For the test cases explored, our results suggest that \XPSI is capable of recovering the true mass and radius within reasonable credible intervals. 
This work also reveals the main vulnerabilities of the analysis: a significant dependence on noise and the presence of multi-modal structure in the posterior surface. 
Noise particularly impacts our sensitivity to the analysis settings and widths of the posterior distributions. 
The multi-modal structure in the posterior suggests that biases could be present if the analysis is unable to exhaustively explore the parameter space. 
Convergence testing, to ensure an adequate coverage of the parameter space and a suitable representation of the posterior distribution,
is one possible solution to these challenges. 
\end{abstract}



\section{Introduction} \label{sec:intro}
Millisecond pulsars are incredibly valuable resources for understanding the behaviour of matter at extreme densities. 
With densities that can reach several times the saturation density in their cores, \acp{NS} are indeed among the densest objects in our Universe \citep{Lattimer12ARNPS,OE17,Baym2018,Tolos2020,YangPiek2020,Hebeler2021}. 
The main scientific goal of the payload \ac{NICER} \citep{Gendreau2016}, installed on the International Space Station, is to
probe matter at these otherwise inaccessible conditions to constrain the \ac{EoS}. 
It targets \acp{MSP} showing X-ray emission with pulsations. 
This pulsating X-ray emission is thought to originate from the heat deposited at the magnetic poles by return currents \citep[see e.g.,][]{RudermanSutherland1975,arons81,HM01}. 
The thermal X-rays\footnote{In this work, thermal emission and X-rays refer to the radiation originated by the finite temperature of elements describing the \ac{NS} surface. } thus generated carry information about the space-time in which the \ac{NS} is embedded.
Meanwhile the relativistic speed of the \ac{NS} surface and the atmospheric beaming break the degeneracy between the effects of the \ac{NS} mass and radius, 
which can then be 
inferred through \ac{PPM} techniques \citep[see][and references therein]{2016RvMP...88b1001W,Watts2019,Bogdanov2019b,Bogdanov2021}.\\

We use the \ac{X-PSI}\footnote{\url{https://github.com/xpsi-group/xpsi}} \citep[][]{xpsi} software package, which is designed to simulate the thermal X-ray emission of \acp{MSP} and estimate the model parameter values that allow for a good representation of a specific data set.
By adopting sampling software like \MultiNest \citep{Feroz2008,Feroz2009, Feroz2019}, and more specifically \PyMultiNest \citep{PyMultiNest}, \XPSI provides a Bayesian inference framework that allows us to explore the parameter space describing the emission model. Model parameters include those that describe 
the temperature patterns on the \ac{NS} surface, observer inclination, distance, interstellar medium, the instrument response and the \ac{NS} mass and radius. 

To establish the reliability of inferences with \ac{PPM}, it is necessary to carry out parameter recovery simulations\footnote{ I.e. simulations aimed at verifying whether our inference processes identify posterior distributions that are statistically consistent with the parameter values injected to build the analysed synthetic data. }, where the analysis pipeline is deployed on synthetic data with known input parameters, to see how well those are recovered. While some parameter recovery simulations using \XPSI have already been reported, \citep{riley_thesis, Bogdanov2019b, Bogdanov2021}, there are other crucial aspects of the analysis process that still need to be explored to establish the robustness of previous and current findings: this is the aim of the current paper.   
In particular we investigate the impact of the Poisson noise present in the data, the analysis settings and the randomness of the sampling, and explore the important role of multi-modal structures in the posterior surface.
We also perform parameter recovery simulations for the more complex surface temperature patterns that were identified as the preferred geometry in \citet[][hereafter R19]{Riley2019}.\\

Parameter estimation in the context of \ac{PPM}, is a high dimensional problem, requiring a large amount of computational resources. 
For this reason, in this paper we only focus on simulated data representing models and parameter vectors that can reproduce \joo X-ray data 
\citep[][]{bogdanov19a}. 
\joo is the first \ac{MSP} whose emission was analysed and for which results were published by the \NICER collaboration (\citealt[][]{Miller2019}, \citetalias{Riley2019}). These publications also present the first mass inferences for an isolated \ac{NS}. Using \XPSI, \citetalias{Riley2019} found that the \NICER data of \joo could be well represented by a \ac{NS} with a radius of $12.74^{+1.14}_{-1.19}\,\mathrm{km}$, a mass of  $1.34^{+0.15}_{-0.16}\,\mathrm{M_\odot}$ and two hot spots on the southern hemisphere \citepalias[][]{Riley2019} \footnote{Uncertainties are approximations of the 16\% and 84\% quantiles in marginal posterior mass (note that posterior mass does not mean posterior of the mass parameter).}. 
The peculiarity of this latter detail, together with the elongated, arc-shape (according to the \XPSI analysis) of one of these hot spots drew a lot of attention among theorists studying magnetic fields on \acp{NS} in general and \acp{MSP} in particular. These temperature patterns indeed imply the presence of a complex magnetic field with multipolar structure, in contrast to the classical picture of a centered dipolar magnetic field \citep[see e.g.,][]{Bilous_2019,Chen2020,Kalapotharakos2021}. 
 These first \NICER results were also recently confirmed by an external group, which also used the openly available \XPSI software to reproduce those initial \joo analyses \citep{Afle23}.

The derived mass of \joo also valorises the role of this pulsar for \ac{EoS} studies. Its relatively standard mass complements the high mass of \joh \citep[][]{Cromartie2020, Fonseca20}, the second \NICER target whose data and results have been published \citep[][]{Riley2021,Miller2021,Salmi2022}.

In Section \ref{sec:method}, we summarise our inference analysis. We lay out our main questions and how we address them in Section \ref{sec:settings}. 
In Section \ref{sec:results} we show our findings and we discuss them in Section \ref{sec:discussion}.  
We conclude with final remarks in Section \ref{sec:conclusion}. 

\section{Methodology: \XPSI upgrades} 
\label{sec:method}
In this work we adopt the same \XPSI framework currently used for \NICER analyses. 
We build on the findings of \citetalias[][]{Riley2019}, by applying an improved \XPSI pipeline to simulated data that mimics a slightly revised \joo \NICER data set. 
Detailed analysis of this revised data set, which is derived from the one presented in \citet{bogdanov19a}, but uses the latest \NICER response matrix, is the main subject of \citet[][submitted]{Vinciguerra2023}.

The aim of these two papers is to set a baseline for the analysis of new, larger \joo \NICER data set that will soon be available. 
In particular, here we reflect on the current analysis protocol, adopted within the \NICER collaboration, and provide a benchmark to consistently interpret future results concerning \joo. 

\subsection{Brief Outline of \XPSI Inference Analysis}
In the following, we briefly outline the main steps of this analysis and the most relevant features 
following recent 
\XPSI developments.

\NICER registers events (which include photons as well as instrumental noise artefacts) characterised by a well-measured time stamp and a specific pulse-invariant (PI) channel. 
Each \NICER PI channel has a nominal energy band which is related to the real energy of incoming photons through the instrument response. 
The events registered by \NICER are then folded over the spin period of the pulsar of interest ($4.87\,\mathrm{ms}$ in the case of \joo) and binned in phases (32 phase bins in past and current \NICER analyses). 
The \NICER data analysed with \XPSI thus take the form of event counts per PI channel and phase bin. 
This data is then compared to simulated data of the same form, through our likelihood function \citepalias[see Section 2.4.3 and Equation 5 of][]{Riley2019}. 

Simulated data are generated by \XPSI, according to the selected model (see Section \ref{subsec:models}) and parameter vector. 
The models that we employ use relativistic ray tracing techniques and describe: (i) the emission patterns on the \ac{NS} surface, how the emitted thermal X-rays interact with (ii) the \ac{NS} atmosphere \citep[using \texttt{NSX},][]{Ho01} and (iii) the surrounding space-time  \citep[assuming the Oblate Schwarzschild plus Doppler approximation][]{Morsink2007}, 
(iv) how they travel through the interstellar medium to the telescope, and (v) how they are registered by the telescope. 
Every model adopted in our analyses has multiple free variables; they include the mass and radius of the pulsar of interest, which impact the observed data through special and general relativistic effects such as lensing, Doppler shifts and aberration \citep[see ][for more details]{Bogdanov2019b}. 
Within \XPSI, parameter estimation is then performed in a Bayesian inference framework, where the parameter space is explored by the sampling algorithm \MultiNest \citep{Feroz2008,Feroz2009,Feroz2019}, specifically \PyMultiNest \citep{Buchner2014}.

\subsection{Updates since R19}
\label{subsec:method_updates}
Since the early publication of the analysis of \joo \NICER data set \citepalias{Riley2019}, \XPSI 
underwent several changes; most of them have already been outlined in \citet{Riley2021}. 
Below we briefly list the most relevant differences compared to the analyses presented in \citetalias{Riley2019} \citep[for more details see][]{Riley2021}.

{\bf \XPSI version}: in this work for simulations and inference analyses we use \XPSI \texttt{v0.7.9} (\texttt{v1.0.0} and \texttt{v2.0.0} to produce the reported corner plots), an updated version of the package used in \citetalias{Riley2019} (\XPSI \texttt{v0.1}). From version  \texttt{v0.6.0}, \XPSI allows multiple rays to come to the telescope from the same point on the \ac{NS} surface, an effect which operates to create multiple images for a small part of the prior compactness space.

{\bf Modeling of the instrument response}: 
    as in \citet{Riley2021} and \citet{Salmi2022}, 
    we no longer include the Crab as part of our modeling of the instrument response, i.e., in Equation 3 of \citetalias{Riley2019}, 
    $\beta_{R19} = 0$
    (here $_{R19}$ indicates parameter definition according to \citetalias{Riley2019}). 
    We instead use a single parameter $\beta\,[\mathrm{kpc^{-2}}] = \alpha D^{-2}$, where $D$ is the distance in $\mathrm{kpc}$ and $\alpha$ is the energy independent scaling factor that multiplies the reference response matrix. In our analysis $\alpha$ is the only parameter shaping the effective instrument response ($\mathcal{R}_{ij} = \alpha \mathcal{R}^{\star}_{ij}$, where $\mathcal{R}$ and $ \mathcal{R}^{\star}$ are respectively the effective and nominal instrument response for the $i^{th}$ channel and the $j^{th}$ energy interval). Our analysis only depends on $\alpha$ and $D$ through their combination $\beta$, hence the choice of sampling the single parameter $\beta$. The prior on $\beta$ has been constructed using two Gaussian distributions truncated at $\pm5\sigma$ for $\alpha$ and $D$, respectively centered at $1$ and $0.325\,\mathrm{kpc}$ with scale parameters $\sigma$ set to $0.1$ and $0.009\,\mathrm{kpc}$.
    
{\bf Priors}: as in \citet{Riley2021} and \citet{Salmi2022}, we adopt isotropic priors (i.e., flat in the cosine) for inclination and colatitudes of the hot spot centers (see Section \ref{subsec:models} for details concerning the model parameters). 

{\bf Settings}: 
the range of the \NICER PI channels has been limited to $[30,300)$, corresponding to nominal energies of $[0.3-3]\,\mathrm{keV}$, compared to the $[25,300)$ range adopted in the analyses of  \citetalias{Riley2019}. 
    The energy range included in the applied response matrix is also slightly altered, following the changes in the instrument response (the upper limit on the energy considered is now $3.715\,\mathrm{keV}$, compared to $3.6\,\mathrm{keV}$ in \citetalias{Riley2019}). 
    Further differences concern settings and definitions of variables specific to the \XPSI pipeline, which are explicitly listed in the \XPSI version of \citetalias{Riley2019} 
    (\url{https://xpsi-group.github.io/xpsi}), such as the resolution setting for light bending $num\_rays$ (now, as in \citealt[][]{Riley2021} and \citealt{Salmi2022}, set to 512, in \citetalias{Riley2019} to 200) 
    \footnote{Previous settings and definitions can still be reproduced, and also generalised, with derived classes that can be set to determine the parameter values of a specific hot spot.}. 

\subsection{\XPSI Models}
\label{subsec:models}
In \XPSI the shape of a hot spot can be modeled by either one or two overlapping spherical caps. 
In the latter case, one of the caps entirely dominates the emission of the overlap region, masking completely the other component.
Each of these caps emits at a uniform temperature. 
If the temperature of the prioritised one is set to match the rest of the star (in this work always assumed to be zero), it will mask part of the emission from the other without contributing to hot spot radiation (for simplicity, hereafter we refer to such a cap as the omitting component and to the correspondent ceding cap as the emitting component). 
In this way we can allow emitting regions with circular, annular and crescent shapes, as well as dual temperatures. 

Each hot spot component is modeled with a number of cells constituting a grid in azimuth and colatitude. 
Despite the discretisation, the emitting area is correctly accounted for by appropriately weighting the edge cells. 
The radiation emerging from the \ac{NS} surface is then modeled with rays generated from these cells. 
Using relativistic ray tracing \citep[we adopt the Oblate Schwarzschild plus Doppler approximation of][]{Morsink2007}, we infer for each of these cells the emission angle required for the ray to reach the observer, given the  phase of rotation (leaf) and the specific location of the cell on the \ac{NS} surface. 
This in turn determines the intensity received by the observer, estimated at different energies, while also accounting for the temperature and surface gravity of the emitting cell, and the interstellar medium. 
Through the instrument response, we then estimate the events registered by \NICER, to which a background component is also added. 

We model the \NICER data set of \joo with the thermal emission generated by two non-overlapping hot spots on the \ac{NS} surface, as assumed in \citetalias{Riley2019}. This is motivated by the two distinct pulses characterising the data set of interest \citepalias[see Figure 1 of ][]{Riley2019}.

\subsubsection{\XPSI Settings}
\XPSI requires us to set specific run parameters; in the analyses presented in this paper, we follow \citet[][]{Riley2021} and \citet{Salmi2022} and (unless otherwise stated) fix:   
the square root of the approximate number of cells per hot spot \texttt{sqrt$\_$num$\_$cells} to 32; 
the square root of the maximum number of cells in the grid describing the hot spot component \texttt{max$\_$sqrt$\_$num$\_$cells} to 64; the phase resolution in the star frame \texttt{num$\_$leaves} to 64; 
and number of energies at which the specific photon flux is calculated \texttt{num$\_$energies} (defined within the likelihood object) to 128\footnote{Visit the documentation page 
\url{https://xpsi-group.github.io/xpsi/hotregion.html} for more details on the parameter definitions.}.
 We refer to runs adopting these settings as {\it high resolution} runs. 
Due to limitation in computational resources, in combination with the different scope of our paper, for the most expensive models we often adopt a {\it low resolution} setting, given by: \texttt{sqrt$\_$num$\_$cells}=18, 
\texttt{max$\_$sqrt$\_$num$\_$cells}=32, \texttt{num$\_$leaves}=32 and 
\texttt{num$\_$energies}=64. 
Comparing results with these two different resolution settings allow us to assess their impact on our results and evaluate whether we could reduce the required computational resources without compromising the inference outcomes.

\subsubsection{Atmosphere and Interstellar Medium Assumptions}

In this work we assume the presence of a fully-ionized \texttt{NSX} hydrogen atmosphere \citep[][]{Ho01,HH09}. To obtain the specific intensity of the radiation field, we interpolate the values registered in a lookup table, where this intensity is precomputed as a function of effective temperature, surface gravity, photon energy and the cosine of emission angle calculated from the surface normal \citepalias[for more details see Section 2.4.1  of][]{Riley2019}.
The methodology is consistent with the setup of \citetalias[][]{Riley2019} and  is mostly motivated by limitation on computational resources \citepalias[for comments over the validity and limitation of this assumption see Section 4.1.1 of ][]{Riley2019}. However, here as in \citet[][]{Riley2021} and \citet{Salmi2022}, we adopt an extended table, including higher values for the surface gravity. \\
The effect of the interstellar medium is modeled and parametrized with the hydrogen column density $N_{\mathrm{H}}$ as in \citetalias[][see in particular Section 2.4.1]{Riley2019}. 

\subsubsection{Model Naming Convention and Parameters}
\label{subsubsec:models}
Within \XPSI it is possible to adopt models with various levels of complexity to match the data. 
To assist the reader, in Figure \ref{fig:models} we provide a schematic representation of our naming convention for emission models \citepalias[see][for more details]{Riley2019}. 
Each hot spot can be characterised by a single temperature (\texttt{ST}) or two temperatures (dual temperature \texttt{DT}). For this paper we will be interested only in single temperature hot spots. 
In the simplest case, the hot spot is described by an emitting spherical cap, simply labelled \texttt{ST}. 
More complicated shapes can be obtained, for a single hot spot, by overlapping two different spherical caps. 
If one of these components masks the other, the hot spot can assume ring-like or crescent-like shapes. We refer to a hot spot, whose masking spherical cap is not constrained in location (except for the overlapping condition), as protruding single temperature \texttt{PST}. 
So far the applications of \XPSI have been limited to modeling the emission of two non-overlapping hot spots, that we label as {\it primary} and {\it secondary} hot spots.  
If the two hot spots describing the emitting surface pattern of our model can assume the same range of shapes, we add: 
\texttt{-S} if all the parameters of the two hot spots are dependent on each other; and \texttt{-U} if they are all independent from each other. 
Otherwise, the two or three  letter acronyms of each hot spots, separated by a plus, are used to label the model.

All of the two hot spot models adopted so far for \NICER analyses include the parameters reported below (parentheses clarify the components in case where two spherical caps are used to describe a hot spot):
\begin{itemize}[noitemsep]
    \item {\bf mass} $M\,[\mathrm{M_\odot}]$: the mass; 
    \item {\bf radius} $R_{\mathrm{eq}}\,[\mathrm{km}]$: the equatorial radius\footnote{As in \citetalias[][]{Riley2019}, we adopt a flat prior in the joint mass and radius parameter space \citepalias[see Prior paragraph, Section of 2.4.1 of ][for more details]{Riley2019} to facilitate subsequent \ac{EoS} analyses \citep[][]{Riley2018}.};
    \item {\bf distance} $D\,[\mathrm{kpc}]$: the distance between the Earth and \joo \footnote{Note that, as mentioned in the introductory part of Section \ref{sec:method}, $\alpha$ and $D$ are not always independently parameterised. \label{note1}};
    \item {\bf inclination} $i\,[\mathrm{rad}]$: the angle between the spin axis and line of sight;
    \item {\bf column density} $N_H\,[\mathrm{cm^{-2}}]$: the neutral hydrogen column density. Following the \texttt{TBabs} model \citep[][updated in 2016]{Wilms2000}, we derive the abundances of all other attenuating gaseous elements, dust, and grains from the value of $N_H$;
    \item {\bf temperature of the (emitting, superseding) primary component} $T_p\,[\mathrm{K}]$;
    \item {\bf temperature of the (emitting, superseding) secondary component} $T_s\,[\mathrm{K}]$;
    \item {\bf radius of the (emitting, superseding) primary component} $\zeta_p\,[\mathrm{rad}]$: the angular opening from the center of the \ac{NS} to the center of the (emitting, superseding) primary spherical cap and its circumference;
    \item {\bf radius of the (emitting, superseding) secondary component} $\zeta_s\,[\mathrm{rad}]$: the angular opening from the center of the \ac{NS} to the center of the (emitting, superseding) secondary spherical cap and its circumference;
    \item {\bf colatitude of the (emitting, superseding) primary component} $\theta_p\, [\mathrm{rad}]$: the angle between the North pole, defined by the spinning direction through the right-hand rule, of the \ac{NS} and the center of the (emitting, superseding) primary spherical cap;
    \item {\bf colatitude of the (emitting, superseding) primary component} $\theta_s\,[\mathrm{rad}]$: the angle between the North pole of the \ac{NS} and the center of the (emitting, superseding) secondary spherical cap;
    \item {\bf primary phase shift} $\phi_p\,[\mathrm{cycles}]$: the phase shift of the center of the primary prioritised component (omitting or emitting) compared to the reference phase set by the data;
    \item {\bf secondary phase shift} $\phi_s\,[\mathrm{cycles}]$: the phase shift of the center of the secondary prioritised component (omitting or emitting) compared to the reference phase set by the data;
    \item {\bf energy-independent scaling factor alpha $\alpha$}: which multiplies the reference instrument response (more on this in what follows) \footref{note1}.
\end{itemize}

In general, our models suffer from many degeneracies \citepalias[see Section 2.5 of][for more details]{Riley2019}.

{
    \begin{figure*}[t!]
    \centering
    \includegraphics[
    width=15cm]{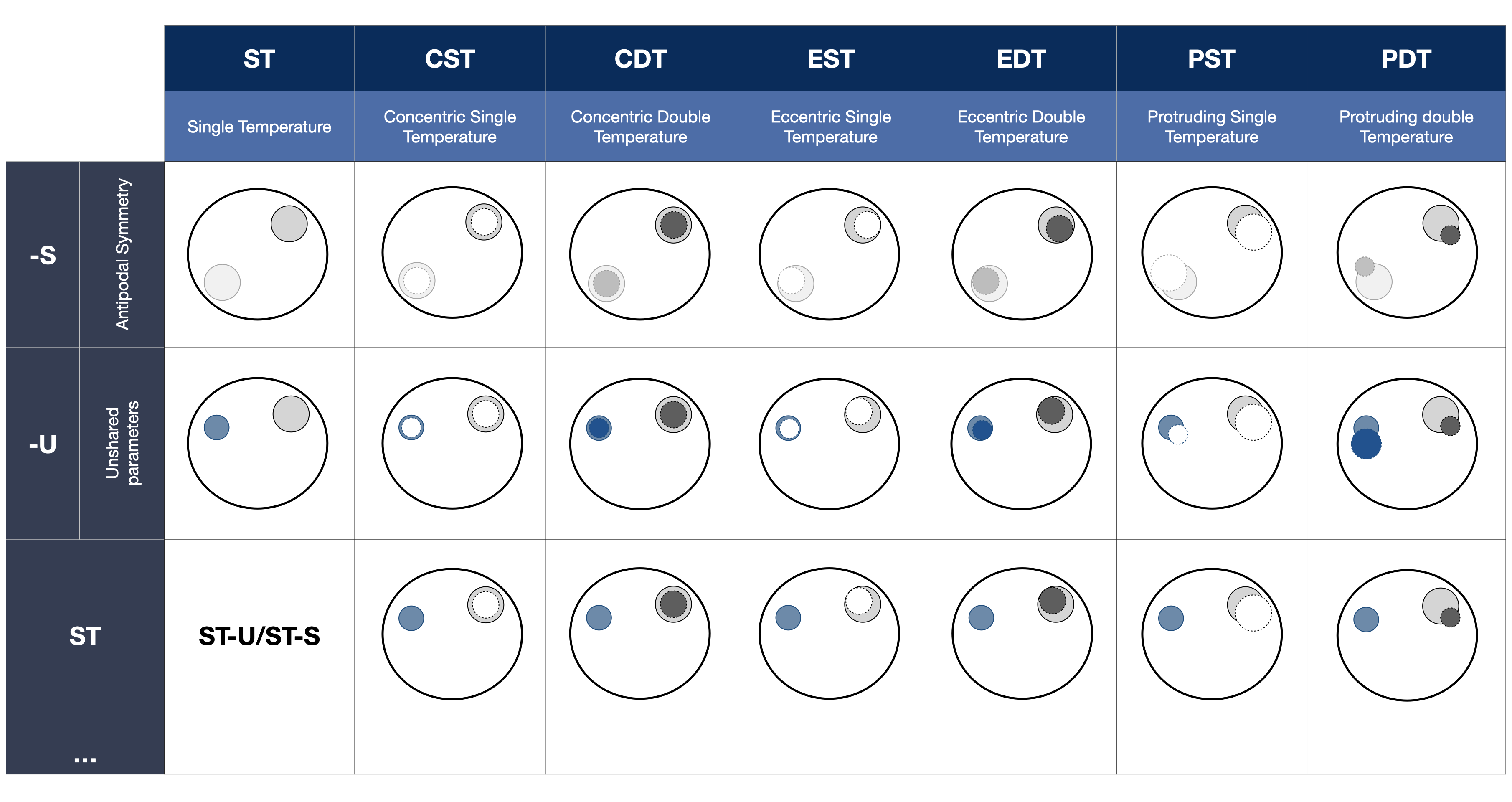}
    \caption{\small{Schematic representation of naming convention adopted within \XPSI. Note that the protruding {\it P} configurations include the eccentric {\it E} ones which, in turn, include the concentric ones {\it C}. In the case of antipodal symmetry ({\it -S} in the table), the lightest hot spot indicates that it is located on the hemisphere opposite to the observer. The dots in the last row of the table suggest how additional models could be built, allowing different geometries for the two hot spots.}
    }
    \label{fig:models}
    \end{figure*}
}

Motivated by the findings in \citetalias{Riley2019}, in this work we apply two different models: \texttt{ST-U} and \texttt{ST+PST}. In \citetalias{Riley2019}, \texttt{ST-U} was disfavoured compared to more complex models in view of their correspondent evidences. 
However this model was not flagged by any anomaly in the residuals (see Section 3, paragraph labeled as {\it Posterior predictive performance in} \citetalias[][]{Riley2019}) and therefore represents the simplest and least computationally demanding model able to reproduce the \joo \NICER data. 
\texttt{ST+PST} was the preferred and one of the most complex models examined in \citetalias{Riley2019}.  
Below we briefly outline changes in the description of the  \texttt{ST+PST} model.

\subsubsection{Changes to \texttt{ST+PST} Model Parametrization}
According to the naming convention explained above, in the \texttt{ST+PST} model the thermal emission from the \ac{NS} surface originates from the radiation of a spherical cap with uniform temperature and a second hot spot, whose shape depends on the parameter values determining the relation between an emitting and a masking spherical cap (see also Figure \ref{fig:STPST}). 
We report parameters as they are defined within the \texttt{ST+PST} model in the \XPSI framework (\citetalias[][]{Riley2019} instead reported derived variables, an alternative description).
As in \citetalias[][]{Riley2019}, we assume the most complex (\texttt{PST}) hot spot to be the secondary. 
In this case, the secondary parameters listed above refer to the emitting component of the secondary hot spot, except for the phase, which instead corresponds to the masking region. 
In addition to the list previously presented, this model requires the definition of the following parameters: 
\begin{itemize}[noitemsep]
    \item {\bf radius of the masking region of the secondary hot spot} $\zeta_{o,s}\,[\mathrm{rad}]$: 
    the angular opening from the center of the NS to the
    center of the masking spherical cap and its circumference;
    \item {\bf colatitude of the masking region of the secondary hot spot} $\theta_{o,s}\,[\mathrm{rad}]$: the angle between the North pole of the \ac{NS} and the center of the masking spherical cap;
    \item {\bf azimuth offset of the secondary hot spot} $\chi_s\,[\mathrm{rad}]$: the offset in azimuth between the emitting and the masking spherical caps of the secondary hot spot (the emitting region is taken as a reference).
\end{itemize}
All the parameters of interest are shown in Figure \ref{fig:STPST}. 
{
    \begin{figure}[t!]
    \centering
    \includegraphics[
    width=8cm]{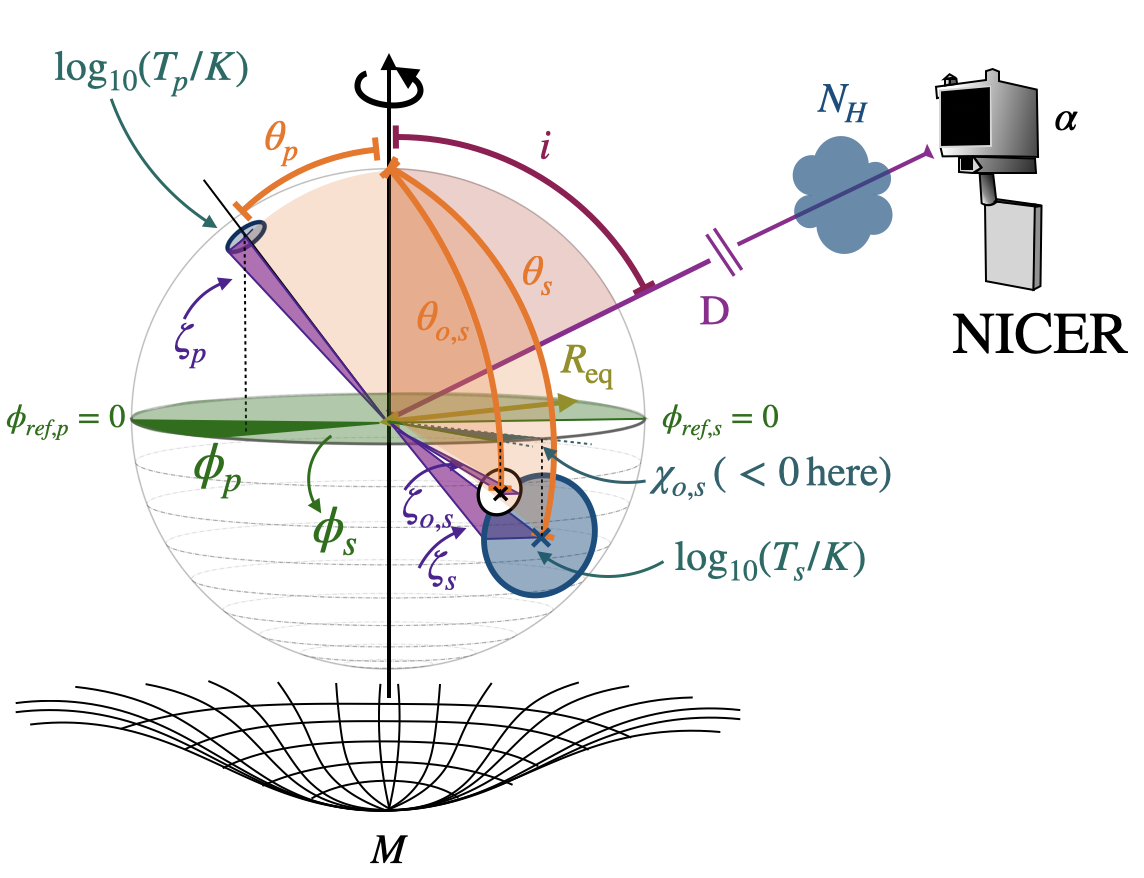}
    \caption{\small{Schematic representation of the \texttt{ST+PST} model and the parameters describing it.}
    }
    \label{fig:STPST}
    \end{figure}
}
Note that despite the change in the reported variables, our inference analyses are based on the same prior parameterisation described in Section 2.5.6 and Appendix B of \citetalias[][]{Riley2019}, except for the following modified rejection rule.

In general in \XPSI, we require that the emitting spherical caps of the two modeled hot spots do not overlap. 
In \citetalias[][]{Riley2019}, the implementation of this condition prevented the primary \texttt{ST} from overlapping also with the omitted part of the emitting spherical cap describing the \texttt{PST} hot spot.  There is however no physical reason to exclude such configurations from consideration. 
Therefore in this work, the primary is allowed to overlap with the secondary masking cap as long as it does not overlap with the non-masked mesh cells of the emitting component, defining a Comprehensive Hot spot 
prior \footnote{Due to the nature of the resulting spherical geometry calculations, the project to change these priors became also known as the {\it Circles of Hell}.}.

\subsection{MultiNest}
\label{subsec:multinest}

In our inference runs we use \MultiNest 
to explore the parameter space. 
Parameter estimation is a by-product of nested sampling algorithms \citep{Skilling2004} as \MultiNest, which target the computation of the evidence. 
Conceptually, to perform such calculation, they start from a number of initial samples (live points) that explore the whole prior space and evolve them to define iso-likelihood contours of higher and higher values, enclosing increasingly smaller prior volumes.  The process continues until the change of evidence, due to the contribution of the remaining, currently-enclosed prior volume, is estimated to be less than 
user-defined threshold, which sets the termination condition. 
Samples are uniformly drawn by 
\MultiNest from a unit hypercube prior volume and are converted to physical parameter values by inverse sampling. 
\XPSI interfaces with \MultiNest through \PyMultiNest \citep{Buchner2014} by defining priors and the likelihood function. 
In our analysis, we employ the same background-marginalised likelihood function for phase-folded and binned events  described in Equation 4 and 5 of \citetalias[][]{Riley2019} \citep[see also][]{Miller2015}. 
To probe our parameter space we inverse sample from our priors, as defined in \citetalias[][]{Riley2019} and at the beginning of Section \ref{subsec:method_updates}.

The use of \MultiNest requires the definition of a range of settings. 
In particular in our standard inference runs, we specify the following parameters which can potentially affect the results of our analyses.
\begin{itemize}[noitemsep]
    \item {\bf sampling efficiency (SE)} $e$ (or equivalently the {\bf expansion factor} $1/e$): this parameter sets the enlargement factor applied to the prior volume adopted during the sampling procedure \citep[Section 5.2 of ][]{Feroz2009}. This parameter is introduced in \MultiNest to widen the prior volume defined by the clusters (ellipsoids), since they may not be optimal in approximating the iso-likelihood contour (suggested values are 0.3 for evidence estimates and 0.8 for parameter estimations). 
    In practice, the value we set in \XPSI is later 
    scaled 
    by the fraction of the unit hypercube sampling space effectively allowed by our prior conditions and rejection rules (see Appendix B Section 5.3 of \citealt{riley_thesis} for details on its implementation in \XPSI);
    \item {\bf evidence tolerance (ET)}: this parameter sets our termination criteria (the suggested value is 0.5) by imposing an upper limit over the contribution of the missing prior volume to the evidence at the current iteration \citepalias[see Appendix A Section 1 of][]{Riley2019}. 
    \item {\bf number of live points (LP)}: this parameter sets how many samples are initially drawn from the prior volume; these are later replaced following the procedure described in \citet{Feroz2009} and schematised in Algorithm 1 of the same reference (in \citealt{Feroz2009} an example is given with 400 LP, and similar values are reported for \texttt{UltraNest} as well, see \citealt{Buchner2021}). 
    \item {\bf multi-modal} or {\bf mode-separation} method ({\bf MM}): when this modality is used, the samples associated with the identified modes are evolved independently and locked to the correspondent mode. The number of live points associated with each mode is determined by the prior mass of each mode upon mode separation. 
\end{itemize}
Accuracy and precision of evidence estimates and posterior distributions increase with low sampling efficiency, low evidence tolerance and high number of live points. 
Whilst making the evidence calculation less efficient, enabling the mode-separation allows us to recover parameters describing disjoint modes identified by \MultiNest. 
The resulting broader understanding of the posterior surface allows us to put the found solutions into a wider context. 
We can compare them against expectations derived from independent inferences and phenomena e.g. other \NICER targets or gravitational wave estimates. 
Unfortunately the computational cost of the analysis also increases with number of live points, low sampling efficiency and low evidence tolerance. Compromises are therefore required. 
Below we explore the impact of differences in \MultiNest settings on the inferred results, while limiting the computational cost.
In particular, we verify the robustness of our inference results 
employing variations of our {\bf reference set up}, defined by 
the same \MultiNest setting configuration adopted in most of the analyses of \citetalias[][]{Riley2019}: {\bf SE 0.3, ET 0.1, LP 1000, MM off}. 
\section{Simulations and Tests: the Case of PSR J0030+0451}
\label{sec:settings}
\subsection{Our Main Lines of Enquiry}
\label{subsec:quests}
This work expands the previous studies reported in \citet{riley_thesis} and \citet{Bogdanov2021}.
In particular we aim to explore the robustness of \XPSI parameter recovery, i.e. checking whether the injected parameter values are recovered within statistically expected credible intervals, for configurations that resemble those emerging both from \citetalias[][]{Riley2019} and a revised 
\joo data set \citep[][submitted]{bogdanov19a, Vinciguerra2023}. For this reason we test:
\begin{itemize}[noitemsep]
    \item different Poisson noise realisations;
    \item different \MultiNest and \XPSI settings;
    \item different initial random conditions in the sampling process;
    \item different models describing the emission pattern, including the never-before-tested and favored, according to \citetalias{Riley2019}, \texttt{ST+PST} model (in particular,
    data sets are generated and analysed with \texttt{ST-U} and \texttt{ST+PST} models);
    \item the effect of a mismatch between the model used to generate and analyse the data sets.
\end{itemize}
Ideally to unveil possible biases, verify the statistical properties of our results and assess their reliability, we would set up large scale simulation studies, 
exhaustively exploring the posterior distributions inferred from 
the analysis of the actual data set \citep[][submitted]{Vinciguerra2023}, 
 similarly to what has been done e.g., in  \citet[][]{Berry2015}. 
Through such studies, we could also confirm the expected dependencies of the inference performances on parameter values 
\citep[][and refs therein]{lomiller13}. 
However there is a considerable mismatch between the computational resources available to us and the resources required to carry out such tests. 
We therefore restrict our study to two simulated expected (i.e., in absence of noise) signals, corresponding to two specific parameter vectors, one per model. 
With this limitation, we used about
$\sim 1.\times10^6$ core-hours on the Dutch national supercomputer Cartesius/Snellius \footnote{\url{https://www.surf.nl/en/dutch-national-supercomputer-snellius}}. 

\subsection{Presentation of Injected Data}
Here we describe the simulated signals that we adopt for the inference analyses presented in this work. 
The simulated data sets can be found in the Zenodo repository \citet[][the Zenodo link will be made public, and the files available, once the publication is accepted
]{SIMUJ0030zenodo}. 
Using the \texttt{ST-U} model, we produce 7 different data sets; all of them rely on the same expected signal and parameter vector, but incorporate different noise realisations. 
These are obtained applying Poisson noise, with different random seeds,  over the expected counts per channel and phase bin (grouped in $270\times 32$ bins), calculated from the applied model, parameter vector and correspondent background. The exact procedure is explained in detail in the \XPSI tutorial (\url{https://xpsi-group.github.io/xpsi/Modeling.html#Synthesis}). 
The expected signal is fixed by the maximum likelihood sample found by a preliminary \texttt{ST-U} inference run (SE 0.3, ET 0.1, LP 10 000, MM on) on the revised 
\NICER data set of \joo analysed in \citet[][submitted]{Vinciguerra2023}. 
The posterior sample sets the values of the 13 model parameters outlined in Section \ref{subsec:models}, which in turn determine the simulated thermal emission of \joo. These are consistent with the parameter posteriors found by \citetalias[][]{Riley2019}. 
The specific parameter values adopted for simulation in this work  are reported in Table \ref{tab:inj_values}
and correspond to the geometric configuration reported in the left panel of Figure \ref{fig:inj}. 

Similarly, we generate three  different data sets adopting the more complex  \texttt{ST+PST} model. 
We limit our tests to three  different Poisson noise realisations,  built in the same way as for the \texttt{ST-U} model, since analysing data sets assuming \texttt{ST+PST} is considerably (up to $\sim 90$ times, for the same \MultiNest and \XPSI settings) more expensive than when using the \texttt{ST-U} model. 
These noise realisations are applied on the expected counts obtained given the 16 values of the model parameters reported in the last column of Table \ref{tab:inj_values} and represented as hot spot geometric configuration in the right panel of Figure \ref{fig:inj}. 
These values describe the maximum likelihood sample of a preliminary low resolution \texttt{ST+PST} run (SE 0.3, ET 0.1, LP 10 000, MM on) from the revised \NICER data set of \joo \citep[][submitted]{Vinciguerra2023}.  
This parameter vector resembles the bulk of solutions found by \citetalias[][]{Riley2019} with the same model. 

For all data sets, we also fix the 270 parameters (one per PI channel) that we use to model the phase-independent background \citep[see Section 2.4.3 of][for more details on background modeling within \XPSI]{Riley2021,Salmi2022}. 
Since the signal is constructed by folding over the counts collected over many rotational cycles, 
this background should account for contributions from cosmic energetic particles, X-ray contamination from the Sun, including optical loading, as well as other X-ray point sources in \NICER's field of view (as their time dependence should wash out over in the folding procedure) \footnote{The phase-independent background, however, cannot capture other sources of emission that couple to \joo's rotational period, i.e., X-rays radiated by \joo via processes other than the thermal emission of the hot spots. In the \NICER X-ray bands so far considered for \ac{PPM}, this contribution is normally assumed to be negligible, with the only possible exception being the thermal emission from the remaining part of the \ac{NS} surface. This is in contrast to accreting and bursting pulsars, which constitute possible targets for future missions such as STROBE-X and eXTP \citep{2016RvMP...88b1001W, dmatter_extp,strobex}, where there may be a contribution from hot spot emission reflected from the disk.
}.
The background is chosen to maximise the likelihood of the \NICER revised data set being produced by the hot spot emission described by the 13 (for data sets constructed using the \texttt{ST-U} model) or 16 (for data sets constructed using the \texttt{ST+PST} model) parameter values of Table \ref{tab:inj_values}. 

To produce synthetic data with \XPSI, we adopt the
{\texttt{synthesise$\_$given$\_$total$\_$count$\_$number}} \XPSI function. 
This calculates a mock data set and its associated exposure time from the values of the model parameters and the number of total counts expected from the source and background.

All the data sets analysed in this work have been generated assuming high resolution in terms of number of cells, leaves and energies (see Section \ref{subsec:models} for more details). 
\begin{deluxetable}{c|c|c}[b]
\tablecaption{{ Injected model parameters}}
\tablehead{\colhead{Parameter} & \colhead{\texttt{ST-U} value} &\colhead{\texttt{ST+PST} value}} 
\startdata
$M\,[\mathrm{M_\odot}]$& 1.13 & 1.33\\
$R_{\mathrm{eq}}\,[\mathrm{km}]$ & 10.20& 13.91 \\
\hline
$\beta \,[\mathrm{kpc^{-2}}]$ & 7.19& 9.25\\
$\cos(i)$ & 0.545 & 0.766\\
$N_H\,[\mathrm{cm^{-2}}]$ & 1.40& 0.98\\
$\log_{10}(T_p/\mathrm{K})$ &6.11 & 6.10\\
$\log_{10}(T_s/\mathrm{K})$ &6.10 & 6.10\\
$\zeta_p \,[\mathrm{rad}]$ &0.15 & 0.08\\
$\zeta_s \,[\mathrm{rad}]$ & 0.32& 0.89\\
$\theta_p \,[\mathrm{rad}]$ & 2.45 & 1.97\\
$\theta_s \,[\mathrm{rad}]$ & 2.75& 2.98\\
$\phi_p \,[\mathrm{cycles}]$ & 0.46 & 0.46\\
$\phi_s \,[\mathrm{cycles}]$ & 0.50 & 0.24\\
$\zeta_{o,s} \,[\mathrm{rad}]$ & -& 0.94\\
$\theta_{o,s} \,[\mathrm{rad}]$ & - & 2.98\\
$\chi_{s} \,[\mathrm{rad}]$ & - & -0.70\\
\enddata
\tablecomments{Parameters are given in the same format adopted to define our models; in particular we express the information concerning inclination and temperature respectively in the form of cosine $\cos(i)$ and logarithms $\log_{10}(T)$. 
The reference phase of $\phi_s$ is half a cycle away from the reference phase used to define $\phi_p$, hence the phase difference between primary and secondary is $\phi_s +0.5 - \phi_p$.
}
\label{tab:inj_values}
\end{deluxetable}
{
    \begin{figure*}[t!]
    \centering
    \includegraphics[
    width=16cm]{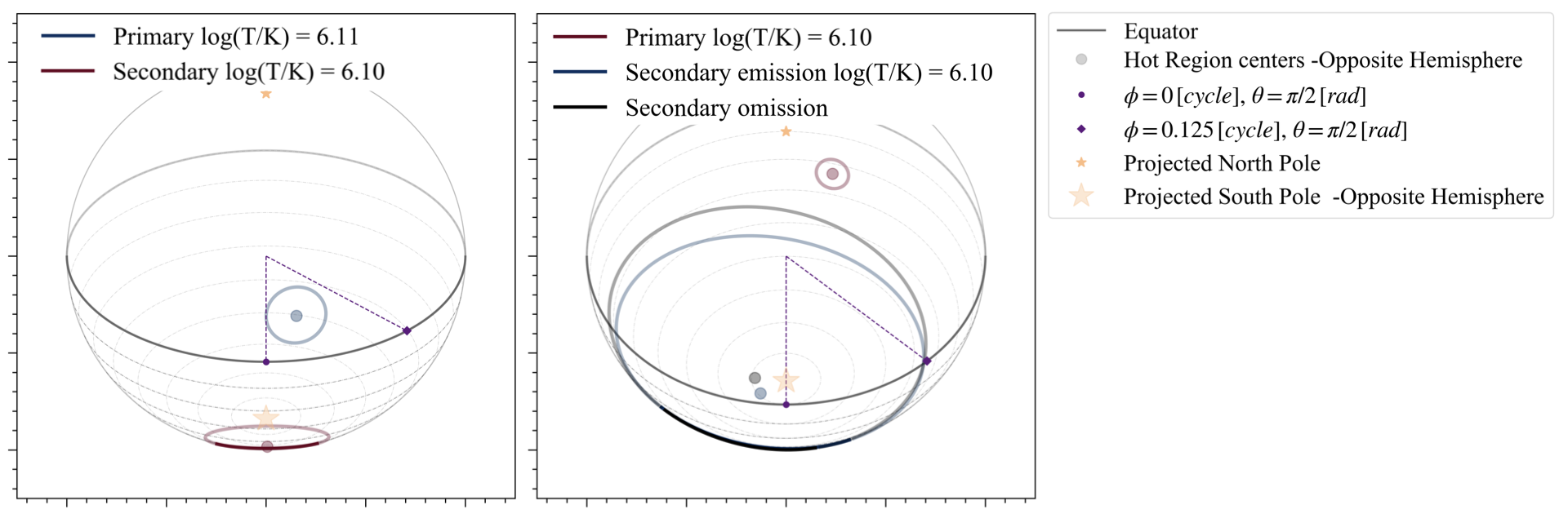}
    \caption{\small{Schematic representation of the geometric configurations, as seen from Earth, of the \ac{NS} hot spots adopted to create the data analysed in this work. The parameter values corresponding to the \texttt{ST-U} and \texttt{ST+PST} models, respectively shown on the {\it left} and {\it right} panels, are reported in Table \ref{tab:inj_values}. The configurations show, in solid lines, the hot spot section visible to us at phase $\phi=\phi_p = 0.0$ cycle (the reference phase of the primary hot spot); in transparency we show the component parts on the hemisphere which, at this rotation phase, is opposite to the observer. With a second point on the equator we also display how the \ac{NS} rotates. The blue/red line is used to mark the hottest/coldest component. We also remind the reader that for the \texttt{ST-U} configuration, the primary hot spot is defined as the component with lower colatitude, while for the \texttt{ST+PST} it is set by the hot spot description as a single spherical cap with uniform temperature.} 
    }
    \label{fig:inj}
    \end{figure*}
}
 
\subsection{Performed Inference Runs}
To investigate the robustness of \XPSI inference analyses, 
we set up a number of inference runs on our simulated data sets. 
Given our limited computational resources and the overall adequacy of the \texttt{ST-U} model in explaining the \joo \NICER data set (see Section \ref{subsubsec:models}), 
we investigate the various performance dependencies listed in Section \ref{subsec:quests}, employing the cheapest \texttt{ST-U} model in the majority of our cases. 
\subsubsection{Inferences with \texttt{ST-U} Models}
\label{subsec:STUtests}
All inference runs performed with the \texttt{ST-U} model are carried out with the high resolution \XPSI settings 
(the same settings used for the data generation) and are reported in Table \ref{tab:STUtests}. 
Below we briefly motivate our \texttt{ST-U} inference runs, in view of the target tests described in Section \ref{subsec:quests}. 

{\it Noise: } To test the effect of different noise realisations in parameter recovery and the width of credible intervals (particularly for the case of mass and radius), we analyse all 7 data sets with the default \MultiNest settings.  

{\it SE, ET and randomness in the sampling process:} 
Of the 7 data sets built with the \texttt{ST-U} model, we use two to test the effect of different values of SE, ET and variability due to the randomness in the sampling process. 
Motivated by the settings suggested by the \MultiNest authors 
\footnote{\url{https://github.com/farhanferoz/MultiNest}} 
and what was adopted in \citetalias[][]{Riley2019}, we test the SE with additional values {\bf SE: 0.1, 0.8}, while keeping ET, LP and MM constant at their default and the ET with additional value {\bf ET: 0.001}, while keeping SE, LP and MM constant at their default. 
We than repeat all these runs, and the one with the default settings, a second time to test variability due to the randomness in the sampling process. 

{\it LP and MM:} For the same two data sets selected for testing SE and ET, we also perform an additional inference run, using 10 000 live points and adopting the mode-separation method (MM on) to increase our prior exploration and learn more about our posterior surfaces. 

{\it Performance when the data set is created with the more complex \texttt{ST+PST} model:}
We would also like to understand the impact of adopting a model in our inference analysis which does not include all the complexity of the true (in this case simulated) system.  This indeed reflects the situation for our normal \NICER analysis, where the models adopted for inference cannot incorporate every detail of the physics describing the actual physical system. However we normally assume that the collected data is not resolved enough for our analyses to be sensitive to the missing physics.   So to test how sensitive we are to the hot spot shapes, we use one of the data sets generated employing the \texttt{ST+PST} model, and test the performance of our inference pipeline when assuming the \texttt{ST-U} model. In this case indeed we know that the model adopted for our inference lacks the complexity used to generate the data.
In particular we would like to check: if mass and radius can be recovered anyway; if the residuals hint at any inadequacy of the model to reproduce the data; if the evidence helps in identifying \texttt{ST+PST} as the best model (for which we also need an inference run with \texttt{ST+PST} as the assumed model, see below); the relation between the recovered and injected geometrical parameters; and how the identified solutions compare to what was found for \texttt{ST-U} in \citetalias[][]{Riley2019}. 
This test can also highlight degeneracies between models and, as a natural consequence, the presence of multi-modal structure in the posterior surface (since we can consider the different hot spot models as nested). 
As shown in Table \ref{tab:STUtests}, we perform 5 inference runs with different \MultiNest settings to check the robustness of our results.

\begin{table}[]
\hspace*{-2cm}\begin{tabular}{l|l|l|l|l|l|l}
\toprule
Data set & SE & ET & LP & MM & N & Core hs\\
\hline
\hline
{Noise 1 }& 0.3 & 0.1&$10^3$& off& 2 & $\sim$900\\
& & & & & & $\sim$1500\\
& 0.1& 0.1& $10^3$& off& 2& $\sim$2900\\
& & & & & & $\sim$1800\\
& 0.8& 0.1& $10^3$& off& 2& $\sim$500\\
& & & & & & $\sim$1000\\
& 0.3& 0.001& $10^3$& off& 2& $\sim$1700\\
& & & & & & $\sim$2000\\
& 0.3& 0.1& $10^4$& on& 1& $\sim$12800\\
\hline
Noise 2 & 0.3& 0.1&$10^3$& off& 2& $\sim$600\\
& & & & & & $\sim$1300\\
& 0.1& 0.1& $10^3$& off& 2& $\sim$1200\\
& & & & & & $\sim$3000\\
& 0.8& 0.1& $10^3$& off& 2& $\sim$700\\
& & & & & & $\sim$800\\
& 0.3& 0.001& $10^3$& off& 2& $\sim$2000\\
& & & & & & $\sim$900\\
& 0.3& 0.1& $10^4$& on& 1& $\sim$13200\\
\hline
Noise 3 & 0.3& 0.1& $10^3$& off& 1& $\sim$1200\\
\hline
Noise 4 & 0.3& 0.1& $10^3$& off& 1& $\sim$1400\\
\hline
Noise 5 & 0.3& 0.1& $10^3$& off& 1& $\sim$700\\
\hline
Noise 6 & 0.3& 0.1& $10^3$& off& 1& $\sim$800\\
\hline 
Noise 7 & 0.3& 0.1& $10^3$& off& 1& $\sim$1000\\
\hline 
\texttt{ST+PST} & 0.3& 0.1& $10^3$& off & 1 & $\sim$1400\\
 (Noise 1) & 0.8& 0.1& $10^3$& off & 1& $\sim$600\\
& 0.1, &0.1& $10^3$& off & 1& $\sim$2900\\
& 0.3, &0.001& $10^3$& off & 1& $\sim$1100\\
& 0.3, &0.1& $10^4$& on & 1& $\sim$13700\\
\hline
\end{tabular}
\caption{Summary of the inference runs performed with the \texttt{ST-U} model. High resolution is always used for number of cells, leaves and energies. The first column shows the synthetic data used for the inference run (horizontal lines separate different data sets). The different noise numbers indicate different noise realisations. SE: sampling efficiency; ET: evidence tolerance; LP: live points and MM (multi-mode): mode-separation modality describe the \MultiNest settings of the inference run (more details can be found in Section \ref{subsec:multinest}). N represents the number of repetitions of a run. Core hs indicates the CPU core hours used to perform the inference run; note that when two identical inference analyses have been performed, the CPU core hours for each run are reported in two separate and consecutive rows.} 
\label{tab:STUtests}
\end{table}

\subsubsection{Inferences with \texttt{ST+PST} Models}
\label{subsec:STPSTtests}
Because of the high computational costs of inference runs employing the \texttt{ST+PST} model, we often use the low resolution \XPSI settings, reducing the number of leaves, cells and energies compared to what was used to produce the various data sets. 
This change also allows us to explore the robustness of our results when adopting more limited resolution.

The settings used for \texttt{ST+PST} inference runs and their motivation resemble what is reported in Section \ref{subsec:STUtests} for \texttt{ST-U} runs and they are summarised in Table \ref{tab:STPSTtests}. 
In addition to the cases presented for \texttt{ST-U} analyses,
here we also check the effect of external constraints on parameter recovery and the width of credible intervals. In particular we set up three  inference runs assuming that there are tight constraints 
on mass and distance (for one run), and mass, distance and inclination for the other two.  
We choose uncertainties compatible with those being used for other \NICER sources, where these constraints are available. 
For these runs, we modify the above-described priors as follows. 

{\bf Mass prior:} We sample the \ac{NS} mass from a normal distribution, centered on an injected value of $M=1.33\,M_\odot$, characterised by standard deviation $\sigma= 0.053\,M_\odot$ 
and truncated at $\pm5\sigma$; 

{\bf Distance prior:} As mentioned at the beginning of Section \ref{subsec:method_updates}, we use information about the distance to define the prior of the $\beta$ parameter. 
Differently from the other analyses (including what was assumed in \citetalias[][]{Riley2019}), for these inference runs we adopt $\sigma =0.0006$\,kpc (instead of $\sigma =0.009$\,kpc).

{\bf Inclination prior:} Finally we tighten the prior on inclination,
using a truncated normal distribution, with center $\arccos(0.766)$ and $\sigma$ set to $0.0001$, on the inclination and inverse sampling the $\cos(i)$ from the cosine of the cumulative distribution of this function.  
\begin{table*}[]
\begin{tabular}{l|l|l|l|l|l|l|l}
Data set & SE & ET & LP & MM & \XPSI settings & constraints & Core hs\\
\hline
\hline
Noise 1 & 0.3& 0.1& $10^3$& off& LR & NO & $\sim$12100\\
 & 0.8& 0.1& $10^3$& off& LR& NO & $\sim$4700\\
  & 0.8& 0.1& $5\times10^3$&off& LR& NO & $\sim$11600 \\
& 0.3& 0.1& $10^4$& on& LR & NO &$\sim$ 55500\\
  & 0.8& 0.1& $10^3$& off& HR& NO & $\sim$14600\\
   & 0.8& 0.1& $6\times10^3$& off& HR & NO & $\sim$103400 \\
   & 0.8& 0.1& $10^3$& off& LR & MD & $\sim$3200\\
   & 0.8&0.1& $10^3$& off& LR & MDI & $\sim$7000\\
& 0.3& 0.1& $10^4$& off& LR & MDI & $\sim$43700\\
\hline 
Noise 2 & 0.3& 0.1& $6\times10^3$& off& LR & NO & $\sim$23000\\
\hline
Noise 3 & 0.3& 0.1& $6\times10^3$& off& LR  & NO & $\sim$35500\\
\hline
\texttt{ST-U} & 0.8& 0.1& $10^3$& off &LR & NO & $\sim$3300\\
 (Noise 1)& 0.3& 0.1& $10^4$& on &LR & NO &  $\sim$79800\\
\hline
\end{tabular}
\caption{
Summary of the inference runs performed with the \texttt{ST+PST} model. The first column shows the synthetic data used for the inference run (horizontal lines separate different data sets). The different noise numbers indicate different noise realisations. SE: sampling efficiency; ET: evidence tolerance; LP: live points and MM: mode-separation (multi-mode) modality describe the \MultiNest settings of the inference run (more details can be found in Section \ref{subsec:multinest}). {\it LR} and {\it HR} respectively correspond to low and high resolution. The  seventh (second-to-last) column describes the parameters on which we applied constrained priors: M stands for mass, D for distance and I for inclination. The CPU hours needed for each run are reported in the last column.}
\label{tab:STPSTtests}
\end{table*}

\section{Results}
\label{sec:results}
In this section, we present the overall results of our inference runs. 
The data and routines (including some examples of modules adopted by \XPSI for inference) necessary to reproduce the posterior distributions presented in this Section are reported in the Zenodo repository \citet[][
the Zenodo link will be made public, and the files available, once the publication is accepted
]{SIMUJ0030zenodo}.

Since the main goal of the \NICER mission is to measure the masses and radii of neutron stars, we particularly focus on the recovery of these parameters. 
In this list of fundamental variables, we also include the compactness, 
the combination of mass and radius to which our analysis is expected to be most sensitive. 
In Figures \ref{fig:STU_n1a2}, \ref{fig:STU_allnoise},  \ref{fig:STUfromSTPST}, \ref{fig:STPST_noise_settings}, \ref{fig:STPSTfromSTU} and \ref{fig:STPST_constraints} 
we therefore report the posterior distributions of mass, radius and compactness obtained by \XPSI, when adopting \MultiNest to sample the parameter space, and
smoothed with \acp{KDE}
\footnote{{\color{purple} \acp{KDE} are applied to the 1D and 2D marginalised posterior distributions found adopting \MultiNest. 
We observe that the total number of samples (in the \texttt{[root].txt}, \url{https://github.com/farhanferoz/MultiNest}), over which we apply the \ac{KDE}, is mostly dependent on the number of live points. In particular the relation between live points and final samples is approximately linear 
($n_{\mathrm{samples}}\approx 30\times n_{\mathrm{LP}}$, 
where $n$ generically symbolizes number).
}}
from GetDist
\footnote{\url{https://getdist.readthedocs.io}}. 
As in \citetalias[][]{Riley2019}, \citet{Riley2021, Salmi2022}, in the 1D posterior plots we highlight the 
area enclosed within the $\sim16\%$ and $\sim 84\%$ quantiles of the 1D marginalised distribution,  while in the 2D plots we show contours for the $\sim 68.3\%$
credible regions; injected values are reported with thin solid black lines. 
In most of our 2D posterior plots, showing compactness versus radius, the \ac{KDE} interpolation introduces an artefact at the boundary of the compactness limit, applied through rejection rules in our prior definition (similar rejection rules and artefacts are also present in \citetalias[][]{Riley2019}, \citealt{Riley2021,Salmi2022})
\footnote{
The presence of this hard boundary formed through rejection rules, and therefore found in the posterior, cannot be easily passed to the \ac{KDE}, which consequently tries to smooth it 
(this is e.g., visible in Figure \ref{fig:STU_allnoise}, where this 2D plot shows the three  contours, defining different credible regions, approaching each other at the bottom and almost delineating a diagonal, while they should resemble the hard cut off that we see e.g., at the bottom of the mass and radius 2D posterior plot). 
Note that similar, non trivial, hard boundaries are also present in the other 2D plots, however most of the time they do not significantly affect our posterior distributions.}.

\subsection{Inferences with the \texttt{ST-U} Model}
We first focus on our \texttt{ST-U} inference runs, whose settings are summarised in Table \ref{tab:STUtests}.  
In particular here we consider parameter estimations on data generated with the same \texttt{ST-U} model (results obtained with mismatching models are reported in Section \ref{subsec:mismatch}).
\subsubsection{Noise and Settings}
\label{subsec:noiseSTU}
Figures \ref{fig:STU_n1a2}, \ref{fig:STU_allnoise} and \ref{fig:STUfromSTPST} 
show that overall mass, radius and compactness are well recovered by our inference runs. 
This is also demonstrated by the P-P plot reported in Figure \ref{fig:pp-plot}. 
The inferred geometry of the hot spots also resembles the correct configuration shown in the left panel of Figure \ref{fig:inj}. 
In particular we find that, with our default \MultiNest settings, the percentage of parameters recovered withing the 1D 68\% credible interval lies within the expected, although indicative, range $\sim (54-84)\%$ 
\footnote{The reported range is indicative as it is calculated under the assumption of independence between the model parameters, which are instead correlated in non-trivial ways. }, for 5 out of the 7 inference runs characterised by different noise realisations. 
This range expresses the uncertainty due to the finite and, for statistical purposes, relatively low number of model parameters. 
The $\sim (54-84)\%$ range is defined by the $\sim 16\%$ and $\sim 84\%$ quantiles of the percent point function of a binomial distribution characterising a sample of size $n = 13$ (number of inferred parameters per run, for the \texttt{ST-U} model) and rate of success $p= 68\%$ (considered credible interval).  
Since we calculated this uncertainty also at the 
68\% level, our findings (i.e., that 5 of 7 runs exhibit parameter recovery within the expected range) are consistent with expectations. 
The two outliers, generated with noise realisation {\it 4} and {\it 5}, recover respectively 12 and 2 of the parameters within the 1D 68\% credible interval. 
Comparing the two panels in Figure \ref{fig:STU_n1a2} and looking at Figure \ref{fig:STU_allnoise}, we notice that a major role is played by the noise realisation. 
In particular, noise seems to have a greater effect than the \MultiNest settings on the precision and accuracy of our results. 
Figure \ref{fig:STU_n1a2} shows, however, that the impact of \MultiNest settings is also somewhat dependent on the noise realisation. Indeed the right panel (where the analysed data was subjected to the noise realisation {\it 2}) shows a larger scatter in the results compered to the left one (where the analysed data was subjected to the noise realisation {\it 1}).  

In Figure \ref{fig:STU_allnoise}, we notice that the injected values of mass and radius intersect the posterior distributions of the data set labeled with noise realisation {\it 5} only at their tails. 
Even in this case however, our analysis is able to identify the correct compactness. All of our runs find the injected value of compactness within the 68\% credible interval of its 1D posterior distribution, the only exceptions being the runs whose noise realisation is labeled with {\it 2}; in these cases the true value lies just outside this boundary.

As mentioned before, the P-P plot of Figure \ref{fig:pp-plot} summarises the findings outlined above, focusing on mass, radius and compactness, for the 7 different noise realisations tested with the \texttt{ST-U} model and whose marginalised posteriors are shown in Figure \ref{fig:STU_allnoise}. 
Both plots show that the mass is always underestimated, however it stays well within the 3 $\sigma$ \citep{Cameron:2011} level. 
Radius and compactness are well recovered, lying most of the time within the 1 $\sigma$ level. 
{
    \begin{figure}[t!]
    \centering
    \includegraphics[
    width=8cm]{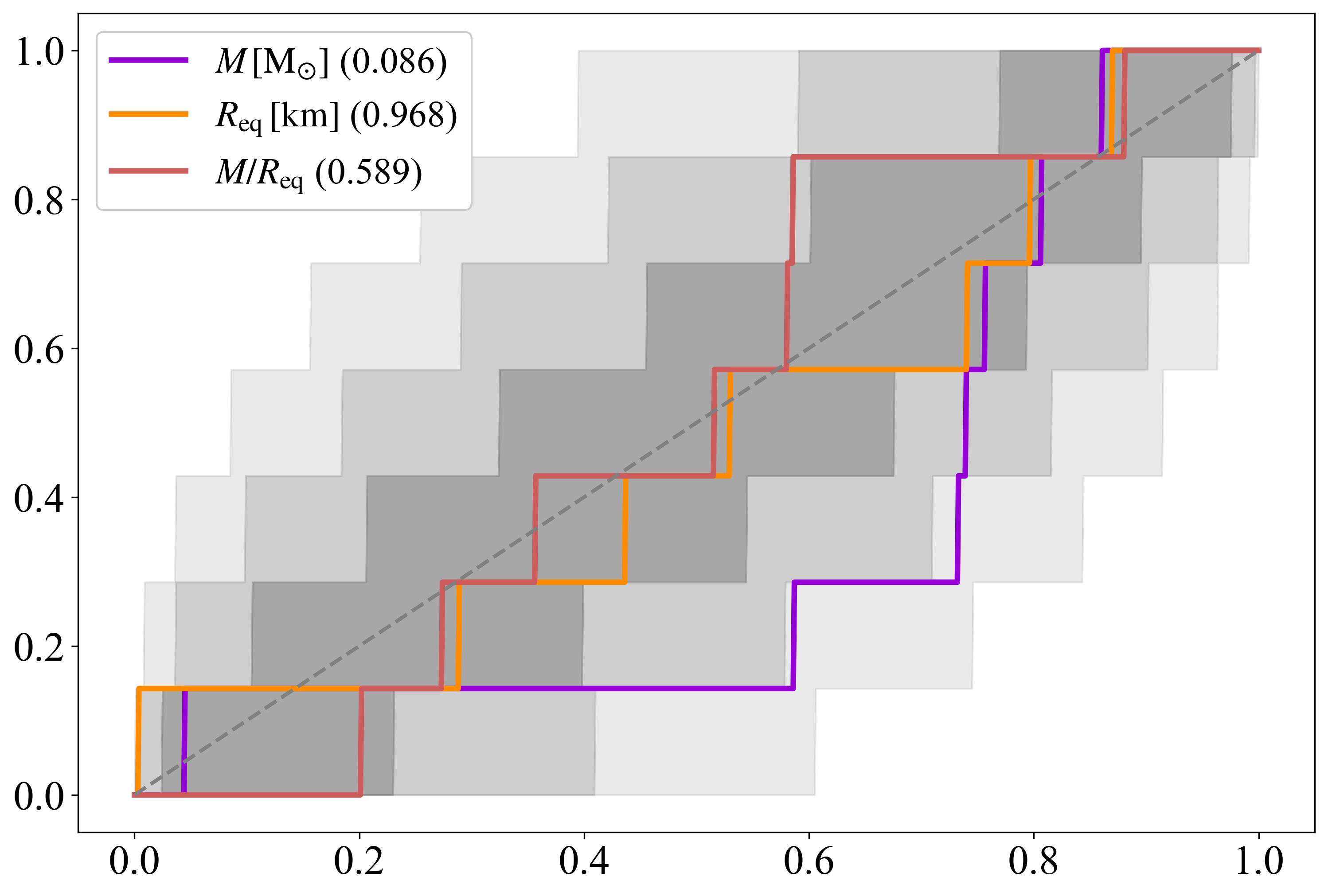}
    \caption{\small{P-P plot for the 7 noise realisations produced and analysed with the \texttt{ST-U} model. 
    It represents, for mass, radius and compactness the cumulative fraction recovered within a credible interval (i.e. the cumulative of the left-side p-value at which the injection is found) as a function of the credible intervals. 
The legend values in parentheses are the p-values from the KS-test against the theoretical uniform expectations (the injected value should appear $p\%$ of the times within the $p\%$ credible interval). 
The grey areas represent the 1, 2, and 3 $\sigma$ confidence intervals on the theoretical expectations, calculated according to \citet{Cameron:2011}. 
    }
    }
    \label{fig:pp-plot}
    \end{figure}
}

These findings corroborate the robustness and reliability of our compactness inferences, at least in absence of unaccounted-for physics. 

\subsubsection{Degeneracies and Posterior Multi-Modal Structure}
As shown in Table \ref{tab:STUtests}, we also run our inference analyses enabling the mode-separation modality. 
Thanks to these runs, we have uncovered a multi-modal structure in our  likelihood and posterior surfaces\footnote{ Given the relatively uninformative priors (for many parameters uniform) that we adopt in our analyses, we expect qualitative one to one correspondence between modes in the likelihood and in the posterior surfaces.}, that were not highlighted in the earlier \citetalias[][]{Riley2019} study\footnote{There is one case that appears to capture an additional mode in the Zenodo repository associated with \citetalias[][]{Riley2019} (\texttt{ST-U} model, inference run {\it 3}).}. 
We find two distinct modes. In terms of hot spot geometries, these two modes are qualitatively similar to the two leftmost plots in Figure \ref{fig:STUfromSTPST_modes}. 
The posteriors of mass, radius and compactness of these two runs, plotted with dashed black lines in Figure \ref{fig:STU_n1a2}, correspond to the main mode. 
In terms of likelihood, there is a clear preference for the main mode; the difference in log-likelihood\footnote{Log-likelihood and log-evidence values are always expressed in natural logarithms.} correspondent to the maximum likelihood samples of these two modes is indeed $\sim 25$. 
Although the secondary modes, found by the two mode-separation runs (respectively on data generated with noise realisations {\it 1} and {\it 2}),  share the main characteristics 
(very low inclination angle, two hot spots similar in size and temperature, almost antipodal in phase, close to the equator and always on the southern hemisphere), they present slightly different properties. 
In particular, the posterior distributions of the \ac{NS} mass and radius have different averages and standard deviations as reported in Table \ref{tab:STUmodes}.

{
    \begin{figure*}[t!]
    \centering
    \includegraphics[
    width=18cm]{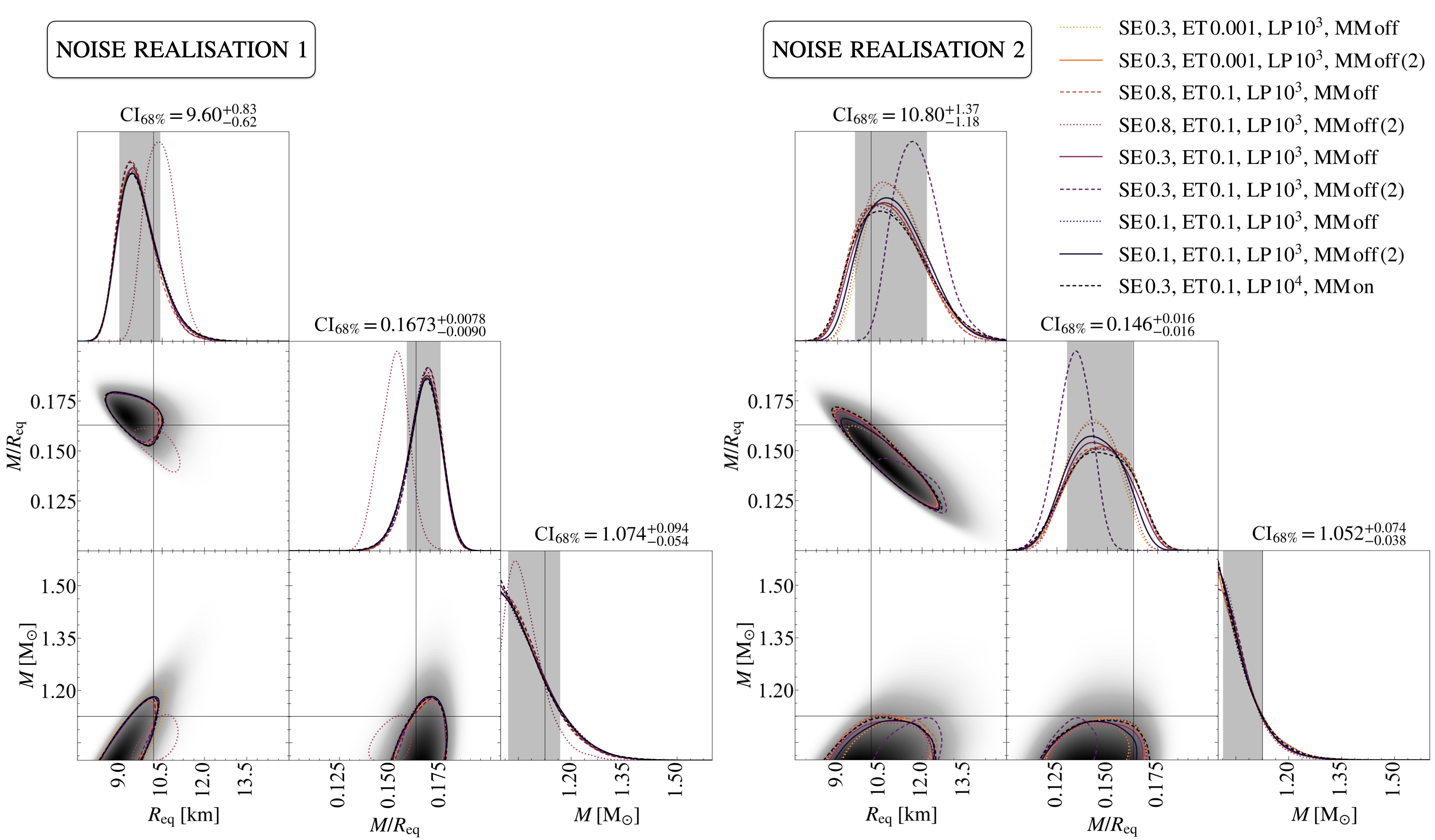}
    \caption{\small{Posterior distributions (smoothed by GetDist \acp{KDE}) from 18 runs, for radius, compactness and mass. 
    Results are obtained using data sets produced with two different noise realisations, labeled as one  (left panel) and two (right panel). The two data sets were generated and analysed with the \texttt{ST-U} model. 
    Each of the two plots shows posteriors from 9 inference runs which use different \MultiNest settings, as reported in the legend (for definitions see Section \ref{subsec:multinest}). 
    On top of the 1D posteriors, we report the 68\% credible intervals  (representing the area within the 16\% and 84\% quantiles in the 1D marginal posterior mass) starting from the median of the distributions. 
    These values, as well as the colored areas, refer 
    to the two runs enabling the mode-separation modality (dashed black lines). 
    The lines in the 2D plots represent the 68\% credible areas of the 2D marginalized posterior, while the shadow refers to the whole distributions of the inferences enabling the mode-separation modality.
    The thin solid black lines represent the injected values. 
    Overall the various settings seem to recover consistent marginalised posteriors, while the different noise realisations adopted to build the two analysed data sets have a significant impact on the shapes of these distributions. 
    The outlier within the runs in the left panel has \MultiNest settings: SE 0.8, ET 0.1, LP $10^3$, MM off (2). 
    The outlier within the runs in the right panel has \MultiNest settings: SE 0.3, ET 0.1, LP $10^3$, MM off (2). 
    } 
    }
    \label{fig:STU_n1a2}
    \end{figure*}
}

{
    \begin{figure}[t]
    \centering
    \includegraphics[
    width=8.5cm]{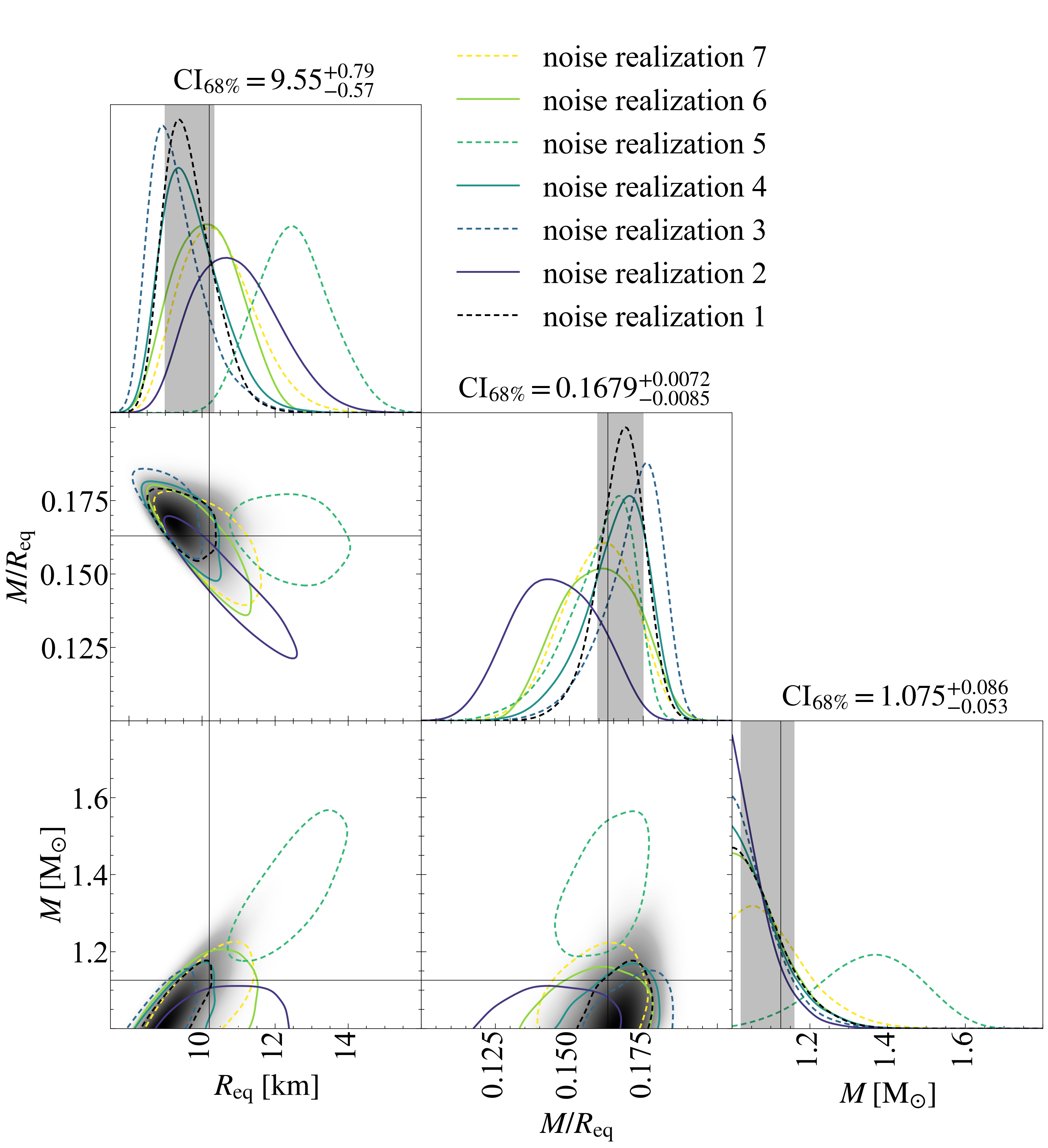}
    \caption{\small{Posterior distributions  (smoothed by GetDist \acp{KDE}) of radius, compactness and mass.  
    Results are obtained using data sets produced with 7 different noise realisations, labeled from 1 to 7 in the legend. The data sets were generated and analysed with the \texttt{ST-U} model. 
    Credible intervals and colored areas 
    refer to the inference run labeled as noise realisation {\it 1} (also represented with dashed black lines). 
    See caption of Figure \ref{fig:STU_n1a2}, for further details on the plot. 
    These inference runs were performed with the \XPSI and \MultiNest settings described in the Sections \ref{subsec:STUtests}, \ref{subsec:multinest} and reported in Table \ref{tab:STUtests}. 
    As expected the different noise realisations introduce some scatter in the marginalised posteriors.
    The outlier (see especially the mass plot) is the run correspondent to noise realisation {\it 5},  the presence of this outlier may be simply due to random fluctuations arising in the sampling process (see indeed also the presence of outliers in Figure \ref{fig:STU_n1a2}). 
    The purple line delineating broad radius and compactness posteriors represents results from  noise realisation {\it 2}. 
    } 
    }
    \label{fig:STU_allnoise}
    \end{figure}
}
 
\begin{table}[]
\hspace*{-1.3cm}\begin{tabular}{l|l|l|l}
&  {\bf Mode 1} &  {\bf Mode 2} &{\bf Mode 3}\\
\hline
$\left<R_{\mathrm{eq}}\right>$\,[km]&9.7,10.9(13.1)& 9.9,13.4(14.6) & (15.3)\\
\hline
$\sigma_{R_{\mathrm{eq}}}$\,[km]&0.7,1.2 (1.5) & 1.3,1.7(1.0)&(0.5)\\
\hline
$\left<M\right>\,\mathrm{[M_\odot]}$&1.1,1.1(1.4)&1.1,1.2(1.5)& (1.6)\\
\hline
$\sigma_M\,\mathrm{[M_\odot]}$&0.1,0.1(0.2)&0.1,0.2(0.2)&(0.2)\\
\hline
\end{tabular}
\caption{Means $\left< \cdot \right>$ and standard deviations $\sigma$ of the mass $M$ and equatorial radius $R_{\mathrm{eq}}$ posterior distributions. The different values correspond to the two (three) modes found by the \XPSI inference run when using the \texttt{ST-U} model to analyse a data set generated with the \texttt{ST-U} model -noise realisation {\it 1}, {\it 2} and, in brackets, with the \texttt{ST+PST}-noise realisation {\it 1}. }
\label{tab:STUmodes}
\end{table}

{
    \begin{figure}[t]
    \centering
    \includegraphics[
    width=8.5cm]{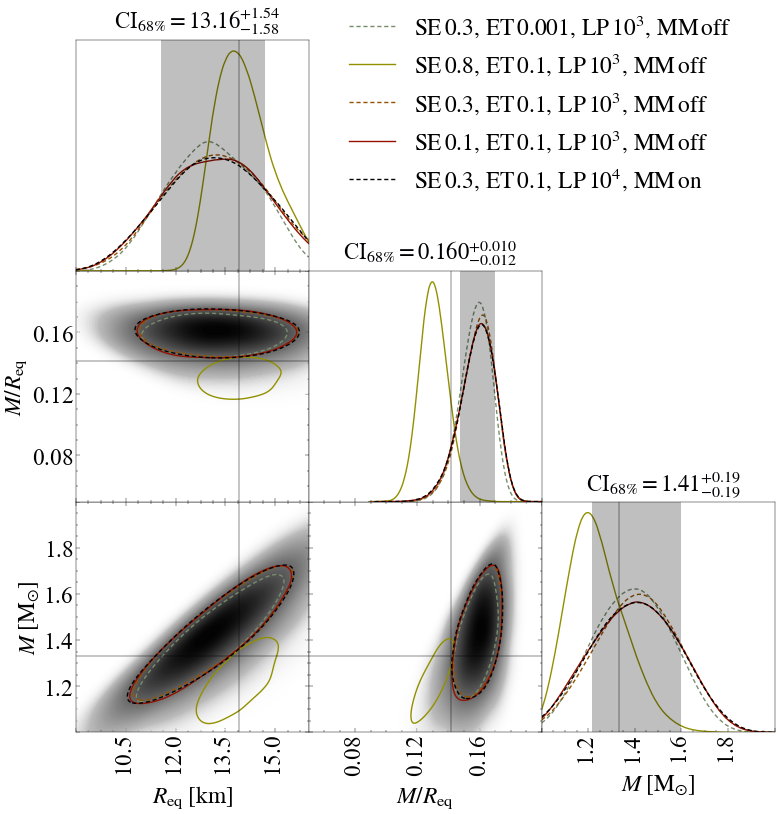}
    \caption{\small{Posterior distributions (smoothed by GetDist \acp{KDE}) of radius, compactness and mass. 
    The data set was generated adopting the \texttt{ST+PST} model (noise realisation label 1) and analysed assuming the \texttt{ST-U} model. 
    Credible intervals and colored areas refer to the inference run enabling the mode-separation modality (also represented with dashed black lines). 
    See caption of Figure \ref{fig:STU_n1a2}, for further details on the plot. 
    The inference runs shown in this plot were performed with the \MultiNest settings reported in the legend (see Section \ref{subsec:multinest} for definitions) and in Table \ref{tab:STUtests}. For all inferences, the injected parameter values are in the bulk of the marginalised posteriors even though the model adopted for inferences does not allow for the complex configuration used to generate the data set. The obtained distributions also well resemble the ones obtained when the correct model is used (see Figure \ref{fig:STPST_noise_settings}). The outlier corresponds to the run with \MultiNest settings: SE 0.8, ET 0.1, LP $10^3$, MM off and again it may simply be due to statistical fluctuations, possibly reflecting the need of more stringent sampling parameters.} 
    }
    \label{fig:STUfromSTPST}
    \end{figure}
}

{
    \begin{figure*}[t!]
    \centering
    \includegraphics[
    width=16cm]{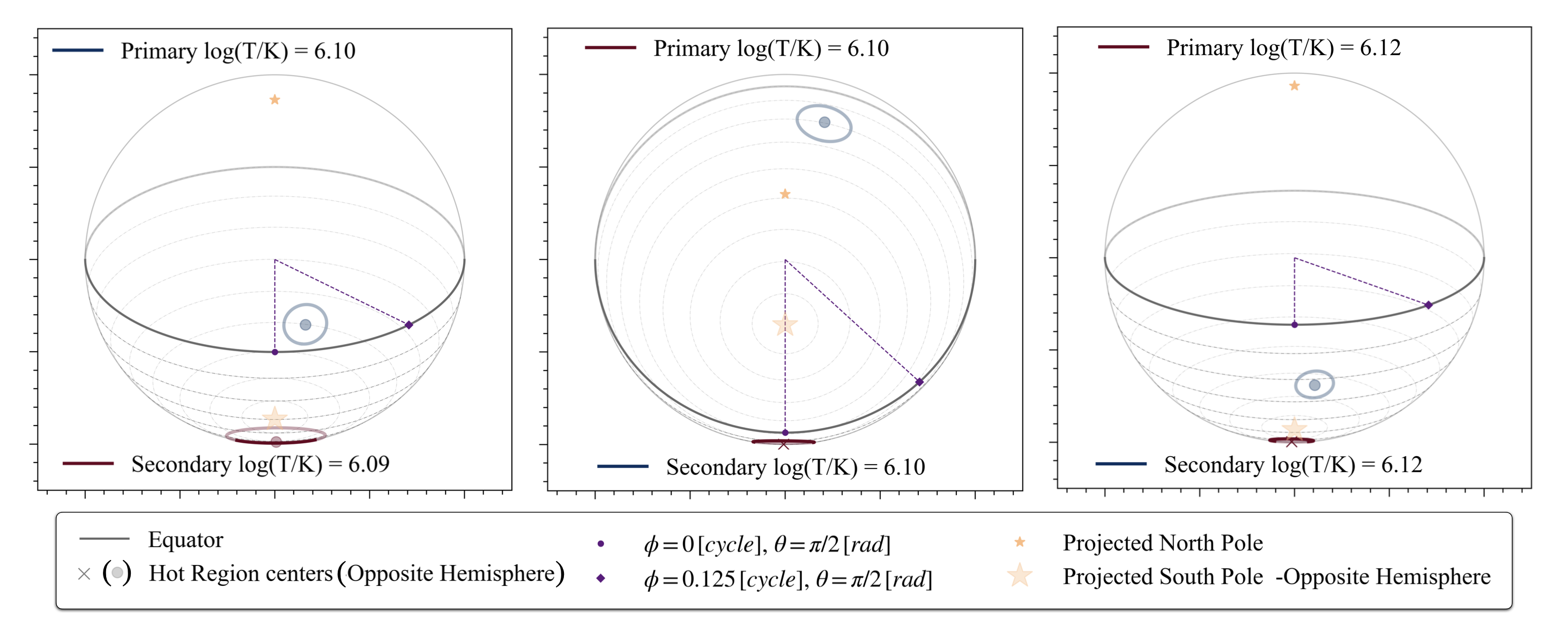}
    \caption{\small{Schematic representation of the the hot spot configurations, as seen from Earth, for the three modes found by the \XPSI inference run when using the \texttt{ST-U} model to analyse a data set generated with \texttt{ST+PST}. The specific configurations correspond to the maximum likelihood sample associated with each mode (for other details, see the caption of Figure \ref{fig:inj}).}
    }
    \label{fig:STUfromSTPST_modes}
    \end{figure*}
}
\subsection{Inferences with the \texttt{ST+PST} Model}
We present here the results obtained with inference runs adopting the \texttt{ST+PST} model, as reported in Table \ref{tab:STPSTtests}, particularly focusing on the analyses of data generated with the same \texttt{ST+PST} model (results obtained with mismatching models are reported in Section \ref{subsec:mismatch}). 
\subsubsection{Noise and Settings}
\label{subsec:ResultsSTPSTnoise}
Figure \ref{fig:STPST_noise_settings} shows the impact of different noise realisations (left corner plot) and different \MultiNest and \XPSI settings (right corner plot) on the inferred posteriors of mass, radius and compactness. Note that for the data set described by the noise realisation {\it 1}, we report results from a run with different \MultiNest settings (LP\,$10^4$ and MM on) compared to the runs on the other two data sets (LP\,$6\times 10^3$ and MM off).
In all the reported runs, the injected values lie within the 2D $\sim 95.4\%$ credible regions. 
Looking at the 1D posterior distributions, we find that, for similar analysis settings, the parameter recovery performance of \XPSI is worse for the more complex model \texttt{ST+PST} than for the simpler \texttt{ST-U} model. 
In particular, when broadening our attention to all the parameters describing the \texttt{ST+PST} model, the three runs reported in the left panel of Figure \ref{fig:STPST_noise_settings} recover within the 1D 68\% credible interval: 7 ($\sim 43\%$, for the case of noise realisation {\it 1}), 5  ($\sim 31\%$, for the case of noise realisation {\it 2}) and 3  ($\sim 19\%$, for the case of noise realisation {\it 3}) parameters over the 16 describing the \texttt{ST+PST} model. 
These recovery rates are all below the expected range of $\sim (56-81)\%$ (calculated as 16\% and 84\% quantiles of a binomial distribution describing a sample of size $n=16$ and success rate $p = 68\%$) and are mostly connected to geometrical parameters. 

The variability due to noise looks comparable to the variability generated by different \MultiNest settings. 
Among them, the number of live points seems to make the biggest difference, in terms of parameter recovery. 
If LP\,$\gtrsim 5\times 10^3$, the posterior distributions become wider and slightly shift toward the correct mass and compactness values. 
Figure \ref{fig:STPST_noise_settings} also demonstrates that,
while noticeably reducing the required computational resources, 
using the \XPSI low resolution described in Section \ref{sec:settings} only slightly modifies our posterior distributions compared to the \XPSI high resolution runs. 

\subsubsection{External Constraints, Degeneracies and Posterior Multi-Modal Structure}
{\it Effects of external Constraints.} 
In Figure \ref{fig:STPST_constraints}, we show the impact on mass, radius and compactness posteriors of different external constraints. 
Comparing it with the results in Figure \ref{fig:STPST_noise_settings}, it is clear that adding constraints on mass and distance significantly reduced the widths of the radius posterior;
however, including 
the constraints on the inclination, in our test case, biases our findings (we discuss these results in details in Section \ref{subsec:constrained_runs}). 

{\it Degeneracies.} 
The complexity of the \texttt{ST+PST} model introduces additional degeneracies between the parameters \citepalias[see also Section 2.5 of ][]{Riley2019}; in particular, in view of our low sensitivity to the smaller details describing the hot spot shapes, many different parameter vectors are able to reproduce quite well the analysed data (see e.g., the small differences reported in Figure \ref{fig:STPST_configuration} and discussed in Section \ref{sec:discussion}).  
This can be qualitatively understood, for example, looking at the top plots of panels A and C, reported in 
Figure \ref{fig:STPST_configuration}. 
They represent the hot spot configurations found in our inference runs on data generated with the {\texttt {ST+PST}} model. 
In particular the top plots of panels A and C represent the maximum likelihood sample of the runs analysing data simulated with noise realisations {\it 1} and {\it 3} (the results for noise realisation {\it 2} mimic the configurations of panel A). 
Although both of these represented configurations can well replicate the simulated data, only the latter recovers a hot spot configuration that resembles the correct one (right panel of Figure \ref{fig:inj}).  This is probably due to the weak sensitivity of our analysis to e.g., the direction of the thermally emitting arc (which indeed faces the right direction in panel C and the wrong one in panel A). 
The additional degeneracies introduced by the complexity of the model therefore compromise the recovery of the model parameters (as demonstrated by the low rate of recovered parameters mentioned in Section \ref{subsec:ResultsSTPSTnoise}), which set the geometry of the emitting \ac{NS} surface.

{\it Posterior Multi-Modal Structure.} When applied to the data set generated using the \texttt{ST+PST} model, our inference runs employing mode-separation modality find two different modes with comparable maximum likelihood values. 
The configuration corresponding to the maximum likelihood samples of these two modes are shown in the top plots of panels A and B,  Figure \ref{fig:STPST_configuration}. 
While the main mode approximately recalls the simulated configuration of the hot spots, the secondary mode resembles the \texttt{ST-U} configuration in Figure \ref{fig:inj}. 
With the averages and standard deviations reported in Table \ref{tab:STPSTfromSTU_modes},
the recovered radius and mass corresponding to this secondary mode are, however, quite close to the injected values.

\begin{table}[]
\begin{tabular}{l|l|l}
 &  {\bf Mode 1} &  {\bf Mode 2} \\
\hline
 $\left<R_{\mathrm{eq}}\right>$\,[km] & 13.8(9.7) & 13.4(9.7) \\
\hline
$\sigma_{R_{\mathrm{eq}}}$\,[km]  & 1.3(0.6) & 1.2(0.7)\\
\hline
$\left<M\right>\,\mathrm{[M_\odot]}$ & 1.4(1.1)  & 1.4(1.1) \\
\hline
$\sigma_M\,\mathrm{[M_\odot]}$ & 0.2(0.1) & 0.2(0.1)\\
\hline 
\end{tabular}
\caption{
Means $\left< \cdot \right>$ and standard deviations $\sigma$ of mass $M$ and equatorial radius $R_{\mathrm{eq}}$ posterior distributions. The different values correspond to the two modes found by the \XPSI inference run when using the \texttt{ST+PST} model to analyse a data set generated with the \texttt{ST+PST} model -noise realisation {\it 1} and, in brackets, with \texttt{ST-U}-noise realisation {\it 1}. 
}
\label{tab:STPSTfromSTU_modes}
\end{table}

{
    \begin{figure*}[t!]
    \centering
    \includegraphics[
    width=18cm]{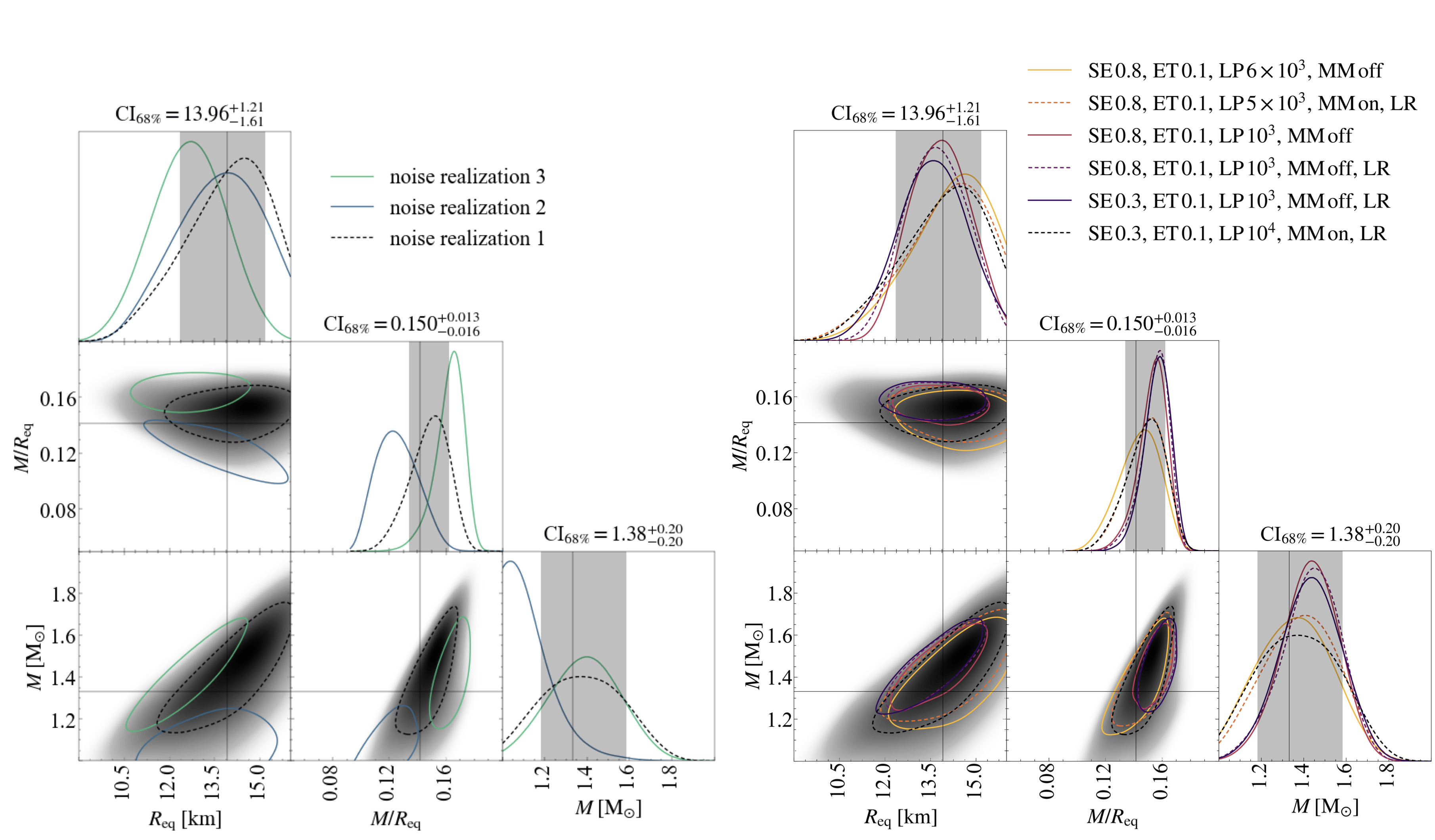}
    \caption{\small{
    Posterior distributions (smoothed by GetDist \acp{KDE}) of radius, compactness and mass. 
    The data sets were generated and analysed adopting the \texttt{ST+PST} model. 
    In the left panel, we present results obtained using data sets produced with three different noise realisations, labeled from one to three in the legend. For all these runs we adopted the \XPSI low resolution setting ({\it LR}), SE 0.3 and ET 0.1. 
     For noise realisation {\it 2} and {\it 3}, we use LP $6\times10^3$ and MM off, for noise realisation {\it 1} LP $10^4$ and MM on (the right panel demonstrates that these two settings lead to similar results). 
     This plot shows that, given these settings, model and observing properties, our recovered posterior distribution is sensitive to the noise realisation adopted to generate the analysed data sets. 
     In the right panel, we show the corner plots corresponding to different runs analysing the data set generated with noise realisation {\it 1} and different \XPSI and \MultiNest settings as shown in the legend. The three curves with broader posteriors represent the runs with $\ge 5\times 10^3$ LP (the first two and the last one in the legend). This corner plot demonstrates the need of a large number of live points to sensibly estimate the width of the marginalised posteriors.   
    All these inference runs are described in Sections \ref{subsec:STPSTtests}, \ref{subsec:multinest} and their details are reported in Table \ref{tab:STPSTtests}. 
    In both plots credible intervals and colored areas 
    refer to the inference run adopting SE 0.3, ET 0.1, LP $10^4$, MM on and LR as \MultiNest and \XPSI settings (represented with dashed black lines), and applied to the data set generated with noise realisation {\it 1}.  
    See caption of Figure \ref{fig:STU_n1a2}, for further details. 
    }
    }
    \label{fig:STPST_noise_settings}
    \end{figure*}
}

{
    \begin{figure}[t!]
    \centering
    \includegraphics[
    width=8.4cm]{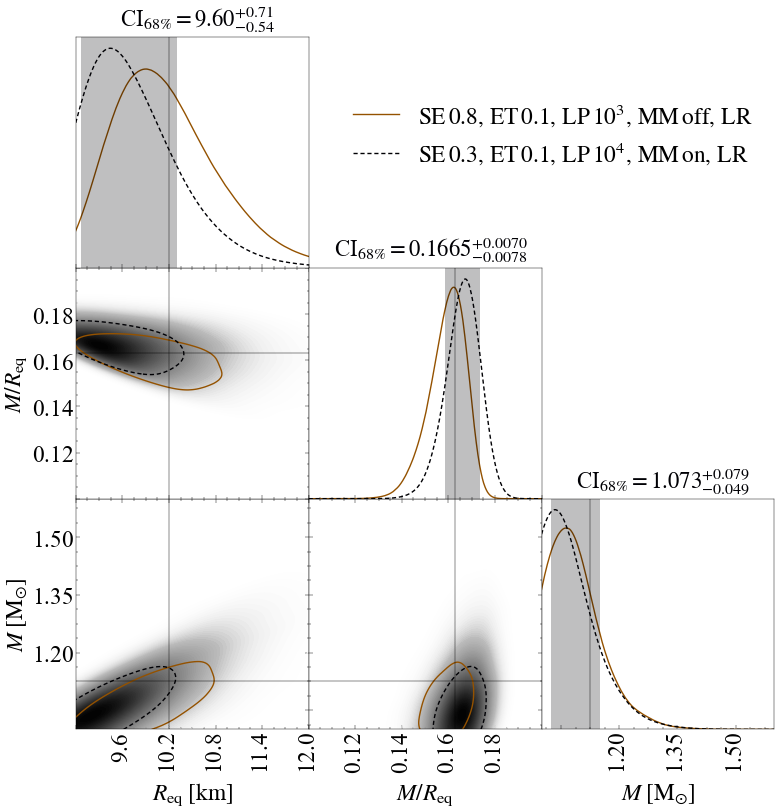}
    \caption{\small{Posterior distributions (smoothed by GetDist \acp{KDE}) of radius, compactness and mass. 
    We present results from two \texttt{ST+PST} inference runs analysing the data set generated with the \texttt{ST-U} model defined by noise realisation {\it 1}; \MultiNest settings are reported in the legend. {\it LR} stands for the \XPSI low resolution setting. 
    Credible intervals and colored areas refer to the mode-separation inference run (also represented with dashed black lines). The injected parameter values are well within the bulk of the obtained marginalised posteriors. These distributions are also similar to ones found when the \texttt{ST-U} model was used to analyse this data set. Increasing the number of live points used in the sampling procedure to $10^4$ slightly shifts the obtained posterior distributions, highlighting that $10^3$ live points are probably not enough to adequately explore the parameter space. 
    See caption of Figure \ref{fig:STU_n1a2}, for further details. 
    }
    }
    \label{fig:STPSTfromSTU}
    \end{figure}
}
{
    \begin{figure}[t!]
    \centering
    \includegraphics[
    width=9cm]{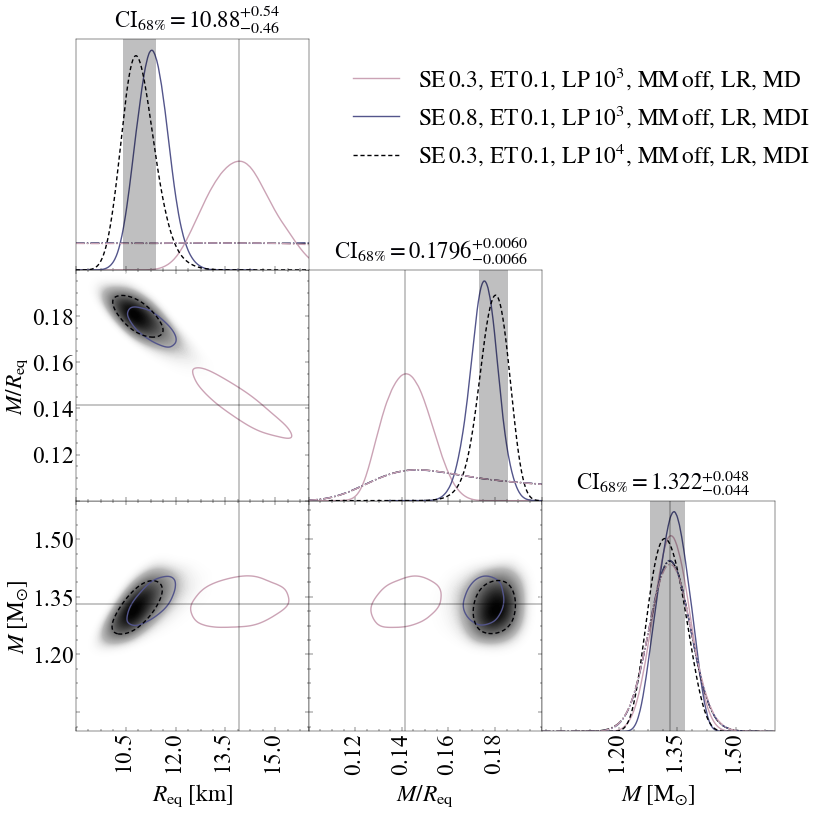}
    \caption{\small{Posterior distributions (smoothed by GetDist \acp{KDE}) of radius, compactness and mass. 
    We report posterior distributions for data analysed and generated with the \texttt{ST+PST} model. With these inference analyses we explore 
    the effect of external constraints on our analysis (see Section \ref{subsec:STPSTtests} for more details). \MultiNest and \XPSI settings, as well as the model parameters a priori constrained (M stands for mass, D for distance and I for inclination), are shown in the legend. 
    For this plot we used the data generated with noise realisation {\it 1}. 
    Credible intervals and colored areas refer to the inference run obtained with constraints on mass, distance and inclination and using $10^4$ live points (also represented with dashed black lines). For clarity, here we also show the 1D marginalised prior distributions on radius, mass and compactness with dash-dot lines. Including the inclination constraints (which is otherwise not well recovered) shifts the inferred marginalised posterior distributions away from the injected values of radius and compactness, highlighting the multi-modal structure and complexity of our posterior surfaces. 
    See caption of Figure \ref{fig:STU_n1a2}, for further details. 
    }
    }
    \label{fig:STPST_constraints}
    \end{figure}
}

{
    \begin{figure*}[t!]
    \centering
    \includegraphics[
    width=17cm]{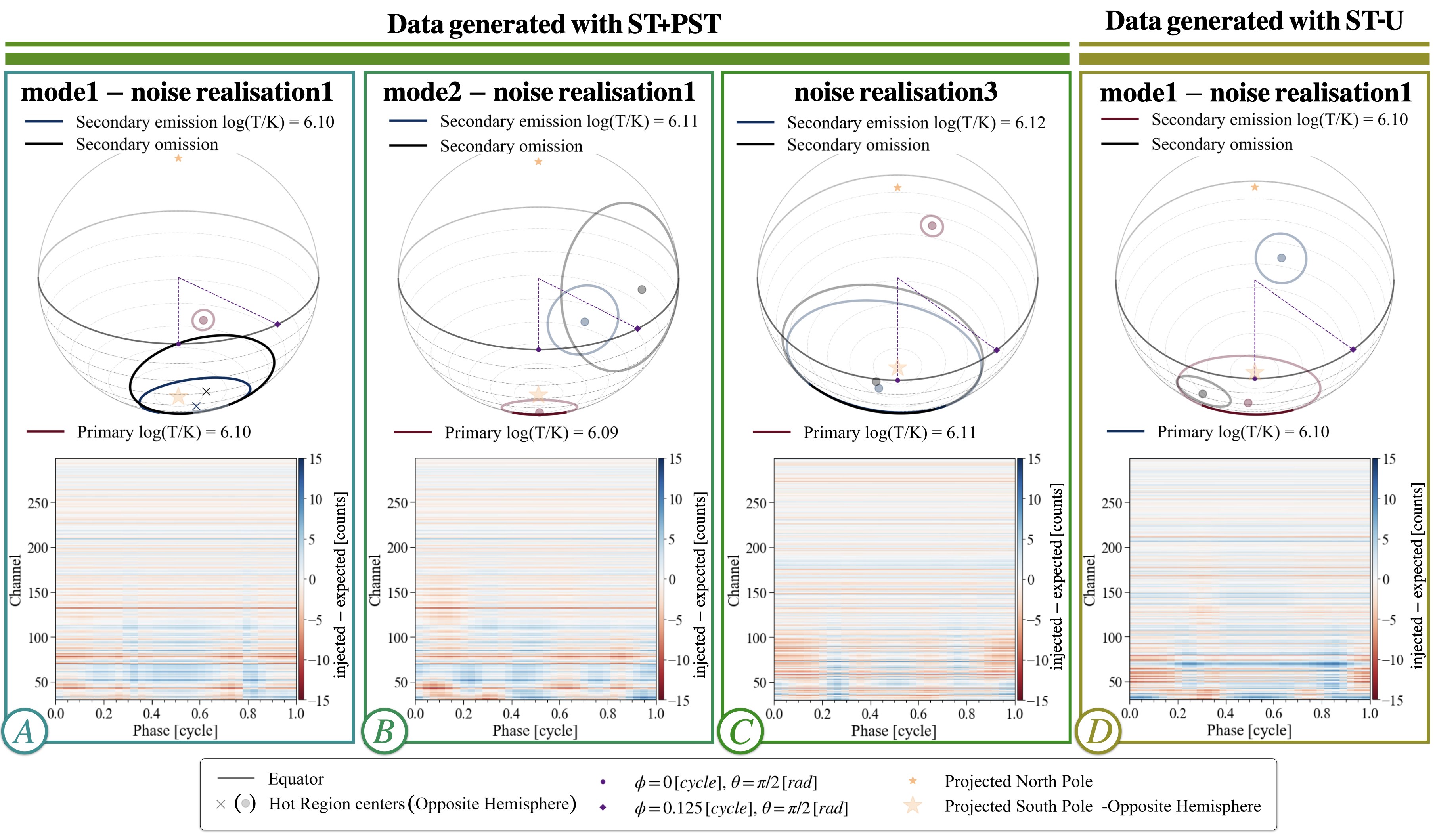}
    \caption{\small{
    Results for three  different \texttt{ST+PST} runs:
    panels A and B refer to the SE 0.3, ET 0.1, LP $10^4$, MM on, LR run on data generated with the \texttt{ST+PST} model and noise realisation {\it 1}; panel C to the  SE 0.3, ET 0.1, LP $6\times 10^3$, MM on, LR  run on data generated with the \texttt{ST+PST} model and noise realisation {\it 3}; and panel D to the SE 0.3, ET 0.1, LP $10^4$, MM on, LR run on data generated with the \texttt{ST-U} model and noise realisation {\it 1}. 
    Mode numbers are specified only for analyses employing the mode-separation modality (also referred to as mode-separation variant). 
    Panel B corresponds to the maximum likelihood sample belonging to the secondary mode; all other panels refer to the maximum likelihood sample of their respective inference runs. 
    {\it Top panels}: schematic representation of the the hot spot configurations, as seen from Earth.
    {\it Bottom panels}: 
    difference in counts between the injected data and the expected counts correspondent to the considered sample of their respective runs (as a reference, in the injected data the maximum count per phase bin and channel is $\sim 700$). The small differences in counts, always smaller then $\sim$ twice the Poisson noise standard deviation, imply that significantly different configurations and parameter values can arise from very similar data sets, assuming the current properties of the \joo \NICER observations. 
    }
    }
    \label{fig:STPST_configuration}
    \end{figure*}
}
 
\begin{table}[]
\begin{tabular}{l|l|l}
{\bf Data/Analysis} &  {\bf \texttt{ST-U}} &  {\bf \texttt{ST+PST}}\\
\hline
 {\bf \texttt{ST-U}} & $-35657.1\pm 0.1$ & $-35655.5\pm 0.1$ \\
\hline
 {\bf \texttt{ST+PST}} & $-35740.5\pm 0.1$ & $-35736.2\pm 0.1$\\
\hline
\end{tabular}
\caption{
Natural logarithm of evidences and their errors (as reported by \MultiNest) for inference runs with settings SE 0.3, ET 0.1, LP $10^4$, MM on. \XPSI low resolution is applied for all inference runs assuming the \texttt{ST+PST} model.
Different rows signify different models used to generate the analysed data; different columns correspond to different models adopted in the inference runs. 
}
\label{tab:evidences}
\end{table}

\begin{table}[]
\begin{tabular}{l|l|l}
 &  {\bf Mode 1} &  {\bf Mode 2}\\
\hline
 $\mathbf{max(\mathcal{L})}$ & $1.14\,M_\odot\,(10.9\,\mathrm{km})$ & $1.01\,M_\odot\,(8.9\,\mathrm{km})$ \\
\hline
 $\mathbf{ max(\mathcal{P})}$ & $1.09\,M_\odot\,(9.0\,\mathrm{km})$  & $1.02\,M_\odot\,(10.1\,\mathrm{km})$\\
\hline
 $\mathbf{mean}$ &$1.12\,M_\odot\,(9.9\,\mathrm{km})$ &$1.09\,M_\odot\,(9.7\,\mathrm{km})$\\
\hline
\end{tabular}
\caption{
Mass and radius values (in brackets) of the maximum likelihood ${max(\mathcal{L})}$, maximum posterior ${max(\mathcal{P})}$ and the mean of the marginalised 1D posterior distributions for the SE 0.3, ET 0.1, LP $10^4$, MM on inference run, employing the \texttt{ST-U} model on a data set generated with the same model and noise realisation {\it 1}. 
}
\label{tab:STU_diff_values}
\end{table}

\subsection{Model Mismatches}
\label{subsec:mismatch}
\subsubsection
{\texttt{ST-U} inferences on data produced with the \texttt{ST+PST} model. }
In Figure \ref{fig:STUfromSTPST} we show 1D and 2D posterior distributions of mass, radius and compactness for \texttt{ST-U} runs on data produced with the \texttt{ST+PST} model. 
This figure suggests that \XPSI inference runs can recover
these parameters 
even when the model used for inference does not capture the full complexity of the ground truth. However, in view of the previous findings concerning our sensitivity to noise realisations, our results cannot be easily generalised, i.e., this could be restricted to a subset of 
parameter values and noise combinations.  
To generalise our findings we would need to consider a statistically significant number of model parameter vectors and noise realisations.
For this data set we also perform an inference run enabling the mode-separation modality. 
We find three modes from this analysis; the configurations corresponding to their respective maximum likelihood samples are reported in Figure \ref{fig:STUfromSTPST_modes}. 
The corresponding means and standard deviations for mass and radius are reported in brackets in Table \ref{tab:STUmodes}.  In this case, the main mode is also clearly dominant in terms of likelihood and evidence calculation, while the other two modes show comparable maximum log-likelihood and local evidences.

So far for \XPSI analyses, we have mostly relied on residuals to verify how well our solution can represent the data. 
In the context of \XPSI, residuals are defined, per bin in channel and phase, as the difference between the data and the inferred expected counts divided by the square root of the same expected counts \citep[see e.g. bottom panel of][]{Riley2019}.
Interestingly, although the \texttt{ST-U} model can not represent a configuration complex as the one injected to simulate the data (shown in the right panel of Figure \ref{fig:inj}), the residuals do not present any anomalous feature and therefore look compatible with Poisson noise.

\subsubsection{\texttt{ST+PST} inferences on data produced with the \texttt{ST-U} model. }
In Figure \ref{fig:STPSTfromSTU}, we report posterior distributions for the mass, radius and compactness obtained when analysing data produced with the \texttt{ST-U} model, assuming the more complex \texttt{ST+PST} model. 
The \texttt{ST+PST} model allows configurations that can well approximate the \texttt{ST-U} ones (\texttt{ST-U} is nested in \texttt{ST+PST}\footnote{The \texttt{ST-U} model can be recovered, within the \texttt{ST+PST} model, setting the angular radius of the \texttt{PST} masking component to zero. In terms of sampling, this value constitutes the edge of the prior of the \texttt{PST} masking component angular radius. }). 
The model can therefore identify, as a main solution, samples that well represent the correct and injected parameter vector. 
Also in this case mass, radius and compactness are well recovered by our analysis. 
In particular, both inference runs on \texttt{ST-U} generated data return 1D/2D posterior distributions whose 68\% credible intervals/regions include the injected values of these parameters. 
However, also here, the various dependencies of our findings and the restricted test cases prevent us from generalising this conclusion.

As for the runs in the right panel of Figure \ref{fig:STU_n1a2}, also here the more computationally expensive \MultiNest settings (LP\,$10^4$, MM\,on) lead to slightly wider and more accurate posteriors compared to the other runs. 
However, now the complexity of the model, and the degeneracies between its parameters, yield two different modes in the posterior, with similar mass and radius (both correctly recovered) and comparable in maximum likelihood and local evidences. 
The corresponding hot spot configurations of the two modes are however significantly different from one another. 
To understand this difference, we can compare the top plots of panels B and D, Figure \ref{fig:STPST_configuration}. 
The configuration corresponding to the maximum likelihood sample of the main mode is indeed represented in the top plot of panel D, Figure \ref{fig:STPST_configuration}. 
The (exact) configuration correspondent to the secondary mode is not reported here, but it is qualitatively equivalent to the secondary mode found analysing data generated with the \texttt{ST+PST} model and shown in the top plot of panel B, Figure \ref{fig:STPST_configuration}. 

Similarly to the previous case, also here the mismatch between the model adopted to create the data set and the one used to analyse it never appears as a clear feature in the residuals. This is, in this case, less surprising, since the model used for inference is the most complex between the two. 

\section{Discussion}\label{sec:discussion}
Here we discuss the results presented in Section \ref{sec:results}.  For the (albeit limited) cases considered in this paper, 
our inference runs on simulated data illustrate the adequacy of \XPSI analysis in recovering mass, radius and compactness given \joo-like \NICER data. 
This reinforces and expands the findings reported in \citet[][]{riley_thesis,Bogdanov2021}, which also included \texttt{ST-U} recovery tests. 
In particular, compactness, mass and radius are recovered within the 95.4\% 1D credible interval (when no additional constraints are applied on the inclination\footnote{As a single pulsar, external constraints on inclination, as well as mass, are not available for \joo. Hence this condition reflects the analysis procedure also followed by the \NICER collaboration.}) for all of the tested data sets, except the one generated with the \texttt{ST-U} model and noise realisation {\it 5}. 
In the following we reflect on the meaning of our findings, particularly focusing on the role of different analysis conditions, and discuss the few anomalous encountered cases and the caveats of our analysis.

The \texttt{ST+PST} inference runs for which we adopted mock constraints on mass, distance and inclination are separated out and discussed in Section \ref{subsec:constrained_runs}.

\subsection{The Effect of Noise, Analysis Settings and Randomness in the Sampling Process}
\label{subsec:noise}
This study shows a clear dependence of our results, including our sensitivity to \MultiNest settings, on the noise realisation. This is shown for the \texttt{ST-U} model in Figures \ref{fig:STU_n1a2} and \ref{fig:STU_allnoise}, and in Figure \ref{fig:STPST_noise_settings} for the more complex \texttt{ST+PST} model. 
This implies that each data set will require its own study to 
assess the robustness of the results.  
In Figure \ref{fig:STU_n1a2} we see indeed that the posterior distributions for the data set created with \texttt{ST-U} and noise realisation {\it 1} (left corner plot) are much more similar to each other than the ones obtained analysing the data set created with noise realisation {\it 2} (right corner plot). 
Note that the posterior distributions in the left corner plot are so insensitive to the different tested \MultiNest settings, that 
even increasing the number of live points by about an order of magnitude \footnote{Our only \texttt{ST-U} run on this data set with LP $10^4$ also enables the mode-separation modality; this effectively reduces the amount of free live points.} does not seem to make any significant difference \citep[despite expectations, see for example][]{bilby2019,Riley2021}.  
However, for both of the \texttt{ST-U} data sets analysed with different \MultiNest settings (i.e. the data generated with noise realisation {\it 1} and {\it 2}), 1 of our 9 inference runs shows a different behavior. 
This is also the case for the \texttt{ST-U} parameter estimation runs for the data set created with the \texttt{ST+PST} model (yellow curve in Figure \ref{fig:STUfromSTPST}).
Given our limited tests, it is not possible to conclusively assess the main origin of such fluctuations. They clearly have a stochastic component, since, for noise realisations {\it 1} and {\it 2}, they appear in only one of two identical analyses; however, it is unclear whether they could be
exacerbated by poorer \MultiNest settings, e.g., by fixing SE to 0.8 (two out of the three  outliers have this setting). 
The poor statistics also prevent a significant evaluation of the 
role played by the noise realisation on the rate of occurrence of these anomalous results. 

Despite the noise fluctuations, compactness is recovered within the $\sim 68\%$ credible interval for almost all cases. 
Exceptions are: the inference run on a data set built with the \texttt{ST-U} model and noise realisation {\it 2} (where the injection value lies just outside it, see right panel of Figure \ref{fig:STU_n1a2}) and the inference run on the data set built with the \texttt{ST+PST} model and noise realisation {\it 3} (which qualitatively recovers the injected hot spot geometry).  
These results are consistent with expectations, although quantitative expectations can only be formulated assuming independence between the parameters. 
Mass and radius are also well recovered by our analyses: 
we recover mass within the $\sim 68\%$ credible interval for 7 of the 10, \texttt{ST-U} and \texttt{ST+PST}, data sets and the radius for 6 of them. These rates both fall within the approximate expected 5-8 range, estimated as explained in Section \ref{subsec:noiseSTU}. 
The main deviation comes from data generated with \texttt{ST-U} model and noise realisation {\it 5}. 
This could either be due directly to the noise realisation, such that repeated inference runs (with the default or better \MultiNest settings) would show the same behaviour, or it could just be due to a random fluctuation (as we see happening for 1 of the 9 \texttt{ST-U} inferences runs on data characterised by noise realisation {\it 1} and {\it 2}). 
We have indeed just argued that the \MultiNest settings required to adequately explore the parameter space may vary for different noise realisations. 
An inspection of this simulated data set does not reveal any particular anomalous feature; we can only identify a slightly lower rate of high counts for channels $\sim (30-60)$ and phases $\sim (0.2-0.6)$ compared to the other noise realisation.
Given the computational resources available to us for this study, we currently cannot fully determine the statistical relevance of this deviation 
nor its origin. Its relatively low rate, however, is in principle consistent with statistical fluctuations and is therefore not particularly worrying.

As shown in Figure \ref{fig:STU_allnoise}, 
different noise realisations can yield very different sizes of the mass, radius and compactness credible regions. 
This finding seems also completely independent from the model adopted to infer the parameter values (see the similarities between the left plot of Figure \ref{fig:STU_n1a2} and Figure \ref{fig:STPSTfromSTU}). 
Our results therefore highlight the crucial role played by stochastic processes on the recovered mass and radius uncertainties and reveal scatter that could complicate and affect their predictions. 

 \subsection{Model Complexity}
Both \texttt{ST-U} and \texttt{ST+PST} models are able to mimic the data of \joo collected by \NICER  \citepalias[see e.g., Figure 1 in][]{Riley2019}. 
Without accounting for noise realisations, the data sets produced, assuming these models and their correspondent parameter vectors as reported in Table \ref{tab:inj_values}, are not only similar in overall counts but also in the hot spot and background  contributions to the data. 
This can be seen in Figure \ref{fig:BKG}, comparing e.g., the mostly overlapping dashed gray and solid black lines, which represent the background counts used (and found in preliminary analyses of the revised \joo \NICER data set \footnote{A similar background was also found in \citetalias[][]{Riley2019}.}) to simulate data with the \texttt{ST-U} and the \texttt{ST+PST} model respectively. 
These strong similarities show that, even for the same background, there are significant degeneracies in the model and parameter space able to explain \joo-like data.
When we use the \texttt{ST-U} model on data produced with the \texttt{ST+PST} model, we find a configuration which very much resembles the one used for generating \texttt{ST-U} data sets and reported in the left panel of Figure \ref{fig:inj}. 
In particular, independently from the model used to create the analysed data set, the \texttt{ST-U} inference run enabling the mode-separation modality finds similar hot spot configurations for the primary and secondary modes. 
When analysing the data set created with the \texttt{ST+PST} model, however, a tertiary mode is also revealed (the geometries of all modes are shown in Figure \ref{fig:STUfromSTPST_modes}).

The \texttt{ST+PST} inference runs show slightly different behaviour: the primary mode found when analysing the data generated with the \texttt{ST-U} model shows a configuration in between the \texttt{ST-U} and the \texttt{ST+PST} one (panel D of Figure \ref{fig:STPST_configuration}). 
Indeed temperatures, inclination and hot spot locations resemble the configuration injected for the \texttt{ST+PST} model, while hot spot sizes and resulting geometries recall the \texttt{ST-U} injection.
Therefore, although the \texttt{ST-U} injected configuration could be very well approximated within the \texttt{ST+PST} model, the larger available parameter space guided the inference process to a geometry which differs from it.  
For two of the three data sets generated with the \texttt{ST+PST} model, we also find a configuration which slightly differ from the injected one. 
Our findings therefore seem to suggest that the complexity introduced by the \texttt{ST+PST} model makes it harder for the sampler to identify the correct parameter values. 
On the other hand, mass, radius and compactness are always well recovered (see Tables \ref{tab:STU_diff_values} and \ref{tab:STPSTfromSTU_modes}); in particular we see that the posterior shapes of these parameters seem to be independent of the model adopted for the analysis. 
This is surprisingly different compared to the situation found in \citetalias[][]{Riley2019}, where the mass and radius changed considerably depending on the model adopted for the \XPSI analysis. 
Differently from the results of  \citetalias[][]{Riley2019} (where the difference in log-evidence between the \texttt{ST-U} and the \texttt{ST+PST} models was of $\sim 10$ units), are also the values of the various evidences. From Table \ref{tab:evidences} we notice that there is never a decisive preference for one model compared to the other, since, given a data set, the evidences differ by just a few units in $\log$. 
Different behaviours compared to the data suggest that our simulations do not capture all features present in the data. 
At this moment, however we cannot conclusively assess if these discrepancies are strictly related to the specific noise realisations (see Section \ref{subsec:noise}), limited to the two considered parameter vectors, or signs of some more profound differences (e.g., some aspect of the physics that is not being modeled). 

{    \begin{figure*}[t!]
    \centering
    \includegraphics[
    width=16cm]{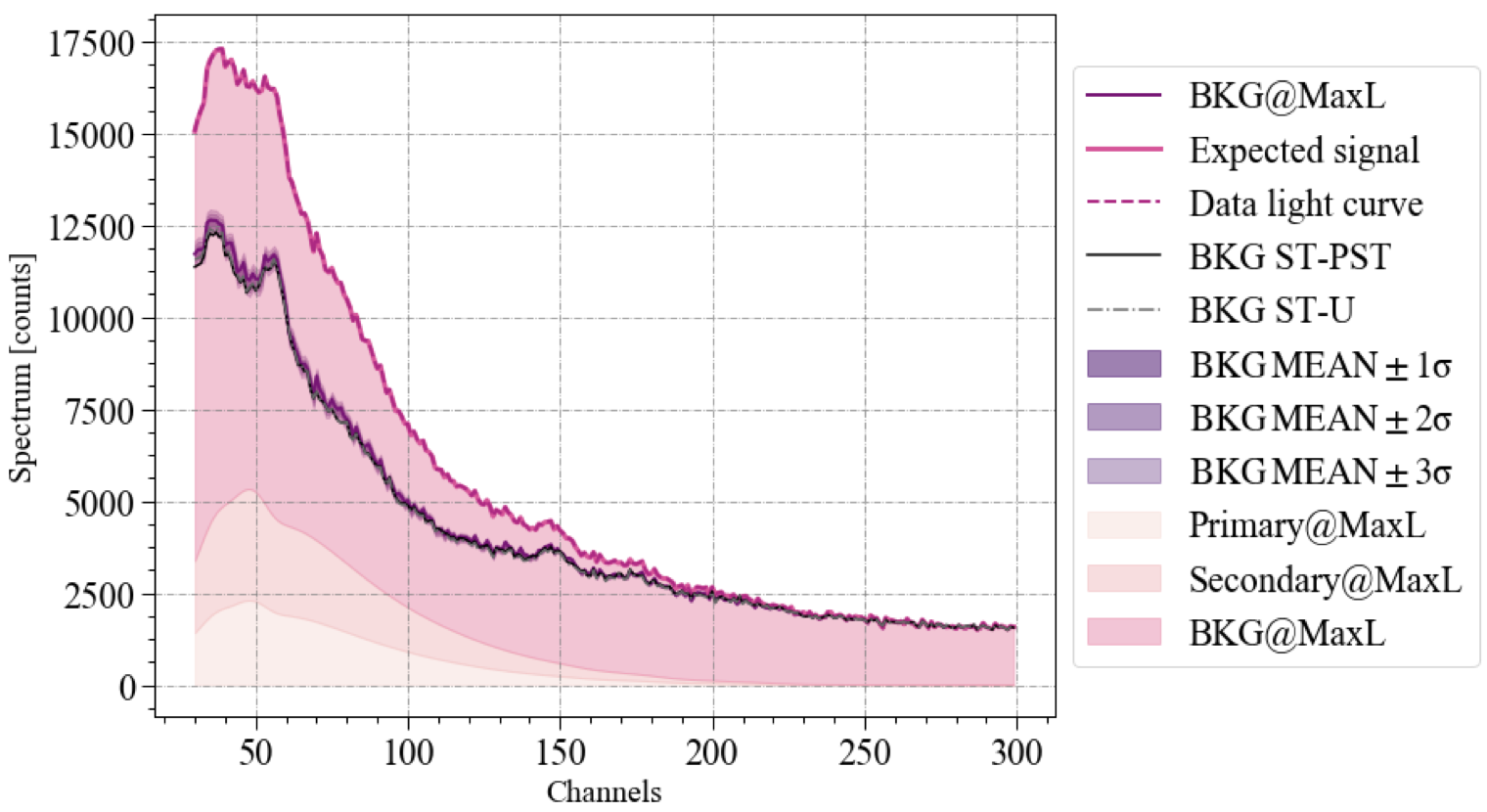}
    \caption{\small{
    Background figure for the \texttt{ST+PST} inference run SE 0.3, ET 0.1, LP $10^4$, MM on, LR on data created with the same model and noise realisation {\it 1}. In shades of pink, from the lightest to the darkest we highlight the contribution to the data in counts per channel of the primary hot spot, the secondary hot spot and the background. 
    The solid pink and dashed fuchsia lines represent respectively the total counts per channel expected according to our model and found in the data.    
    With the solid purple line we show the background correspondent to the maximum likelihood sample of the inference run. 
    From the strongest to the dimmest purple regions we show the 
    $\pm1$, $\pm2$ and $\pm3$ 
    standard deviation regions. 
    The solid black and the dashed gray lines (mostly overlapping) show the background added to the hot spot contribution to obtain the simulated data with the \texttt{ST+PST} and \texttt{ST-U} models respectively. 
    }
    }
    \label{fig:BKG}
    \end{figure*}
}
 \subsubsection{Degeneracies and Multi-Modal Structure in the Posterior and Likelihood Surfaces} 
 \label{subsec:degeneracies}
 A general discussion of degeneracies between model parameters is found in \citetalias[Sections 2.5.4 and 2.5.6 of ][]{Riley2019}; here we comment on them in relation to the specific findings of this paper. In the context of mock \joo \NICER data, {\it our inference runs demonstrate the degeneracies between model parameters via the presence of a multi-modal structure in the posterior surface. } 
  
In this work we took advantage of the mode-separation modality offered by \MultiNest.
 This has highlighted the presence of a multi-modal structure in the posterior surface, which does not comes as a surprise given the different configurations found in the nested models explored for \joo \NICER data in \citetalias[][]{Riley2019}. 
 As we comment below, naturally the extent to which degeneracies populate the parameter space is correlated with the degree of multi-modality present in the posterior surface. 
 This should be kept in mind when comparing evidences between models; indeed higher evidences could arise from the introduction of a more adequate model to describe the data (i.e. for the presence of higher likelihood points) as well as from larger portions of the parameter space rendering similarly good solutions to represent the data.

 For the \texttt{ST-U} inference runs, the difference in likelihood and evidence between the various modes is large enough to strongly prefer the correct mode; 
the performance of \XPSI in recovering injected parameters mimicking the secondary mode, has however not been checked. 
Although the mass and radius of the primary mode are always in reasonable agreement with the injected values, 
Table \ref{tab:STU_diff_values} shows that the 
radius values associated with the secondary mode change considerably depending on the specific considered data set (and therefore noise realisation). 
This variability may be due to an inadequate number of live points covering the specific mode, or due to random fluctuations. 

Looking instead at the \texttt{ST+PST} inference runs 
we find a different situation. 
As mentioned above, in two of the three analysed data sets, we are unable to find the injected geometry (see Figure \ref{fig:STPST_configuration}), even though all runs and both of the flagged modes display mass and radius posteriors compatible with the injected values (see Table \ref{tab:STPSTfromSTU_modes} and Figure \ref{fig:STPST_noise_settings}). 
Indeed multiple hot spot configurations can give rise to very similar 
\joo-like data sets.  
For all three runs in the left panel of Figure \ref{fig:STPST_noise_settings}, the injected configuration had a likelihood difference from the maximum likelihood solution of only a few units in $\log$. 
This can also be understood e.g., by looking at the bottom plots of Figure \ref{fig:STPST_configuration}. These plots represent the difference in counts, per energy channel and phase bin, between the injected data sets 
and the expected one, given by the maximum likelihood sample of that specific run or mode (corresponding to the hot spot geometry represented on the above panel). 
Note that the largest differences occur where the typical counts per energy channel and phase bin are a few hundred, so that the relative difference is never more than a few \% ($\lesssim 10\%$). Given the number of counts characterising these bins, this percentage is always smaller than $\sim$ twice the Poisson noise standard deviation. 
 This means that, assuming the same properties of the revised \joo \NICER data set \citep[for more details see ][submitted]{Vinciguerra2023}, 
  we expect no significant difference between the data produced with the various configurations (whose geometry is represented on the corresponding top panel). 

If we integrate these plots over phase bins $i$ and energy channels $j$, we can 
define the variable $$\mathfrak{D} = \frac{\sum_{i} \sum_{j} |d_{i,j} - c_{i,j}|}{\sum_{i} \sum_{j } d_{i,j}}\,,$$ where $c$ and $d$ represent numbers of counts respectively for inferred sample solutions and the injected data. 
For all the 4 cases (from panel A to D) presented in Figure \ref{fig:STPST_configuration}, we find that the integrated difference $\mathfrak{D}$ between the injected data $d$ and the expected counts predicted by the run or mode $c$ (assuming its maximum likelihood sample) is smaller than the difference between the simulated data in presence and absence of noise $\mathfrak{D_{sim}}\sim 0.056$.
This highlights the presence of some major degeneracies between our model parameters, as introduced in Section \ref{subsec:models}, for a \joo-like data set. 
We can use the top panels of Figure \ref{fig:STPST_configuration} to motivate some of them.
The similar values of likelihoods and evidences between all of these configurations tells us that, with these simulated data sets, we are not very sensitive to the details of the shapes of either hot spot. 
For example the top plots of panels A and C show the arc of the  \texttt{PST} region oriented in opposite directions, and in both cases a visual inspection of the residuals does not highlight any anomaly. 
Similar pulses can therefore be generated even when the parameters describing the hot spot significantly differ (e.g., a difference in the arc direction is rendered with the centre coordinates of the spherical caps having considerably different values). 
Similarly the emission from the \texttt{ST} hot spot seems to be captured by both a circular hot spot as well as an arc, comparing the top plots of panels A and B. 
Moreover we find that, in general, the most likely configurations presented in this paper cluster around values of inclinations between $i \sim 40^\circ$ and $i \sim 60^\circ$; the limits of this range also roughly correspond to the inclinations used to simulate data respectively with the \texttt{ST+PST} and the \texttt{ST-U} model. 
Focusing on the \texttt{ST+PST} inference results, Figure \ref{fig:STPST_configuration} shows that both inclination values can be recovered, independently from the model used to generate the analysed data. 
To generate data comparable to the analysed one, the hot spot geometry needs to adapt to the different inclination values. 
When we have lower inclination values, the hot spot, closer to the equator, needs to have lower colatitude to still be visible to an observer. Similarly the emitting region located closer to the South Pole needs to reach lower colatitude and cover a larger area to still be detectable in the correct phase interval.

 {\it The noise shifts the peak of the likelihood away from the true parameter values} {\it(as expected)} {\it and the sampler does not always identify modes of comparable likelihood}\footnote{Sometimes the main solution found in our inference process significantly differs from the injected one. By calculating the likelihood of the injected parameter vector and inspecting the final posterior samples selected by \MultiNest, it is possible evaluate if a mode has been accounted for or not. 
 Sometimes these investigations lead us to conclude that not all the modes with significant likelihood values have been considered by the sampler. It is however possible for the prior volumes of these modes to be considerably lower than the identified mode. This could, in principle, lead to a substantially low impact of this solution on the evidence, whose estimate is the primary goal of \MultiNest. However this is something that cannot be guaranteed without likelihood evaluations of the corresponding portion of the parameter space. }. 
 
 The \texttt{ST+PST} analyses, for the data sets labeled with noise realisation {\it 1} and {\it 2}, were unable to identify the likelihood peak corresponding to the true hot spot configuration, despite them having comparable likelihood values to the best fitting samples found. 
The absence of configurations similar to the injected one in the posteriors, despite the comparable likelihood value, reveal the inadequacy of the \XPSI and/or \MultiNest settings adopted in our analyses for these specific cases. 
Indeed comprehensive tests, assessing the robustness of the obtained results and the level of coverage of the parameter space for \texttt{ST+PST} inference runs, are computationally demanding and we therefore decided to prioritise preserving compute time to carry out these kind of studies for the analysis of the upcoming and future new data sets.

The inference run on data with noise realisation {\it 3}  (the one which recovered the injected geometry) instead collected samples also resembling the configuration found as the main mode for the other two noise realisations. 
Despite the difference of only a few units in $\log$ likelihood, however, this latter configuration was not prominent enough to form a clear feature in the posteriors.

Importantly none of the solutions found, including the ones pointing to a slightly different geometry compared to the true ones, exhibit any anomaly in the residuals. 
Once we are assured that the parameter space has been exhaustively explored and if multiple solutions are revealed, it is possible to evaluate them considering a broader context, including e.g., radii inferences from other \NICER sources, constraints/indications coming from independent phenomena, such as gravitational waves \citep[see e.g., ][]{Raaijmakers2021}, or even from theoretical advancements. 
Alternatively, this independent information could also be incorporated in follow up test runs with the application of tighter priors on the radius. 

\subsection{External Constraints} 
 \label{subsec:constrained_runs}
The impact of the multiple modes arising from the posterior surface could be, at least in principle, mitigated by external constraints, e.g., on mass, distance and inclination (coming from radio observations) or on the background spectrum. 
Applying such constraints can also considerably reduce the uncertainties on the inferred parameters, including radius. This is clearly visible comparing the sizes of the posteriors in Figure \ref{fig:STPST_constraints} to the ones in Figure \ref{fig:STPST_noise_settings}.

In the cases analysed in this paper, however, tight constraints on inclination end up biasing our results, even affecting the radius inferences, which were otherwise correctly estimated.   
In addition, these biased solutions do not exhibit any anomaly in the residuals. 
Comparing the two {\texttt{ST+PST}} runs with constrained inclination prior, we notice that increasing the \MultiNest resolution settings (in particular increasing the live points and lowering the sampling efficiency) improves the performances of our analysis. 
In particular it increases the likelihood of the maximum likelihood sample by a factor of $\sim 15$ in $\log$. 
The reason becomes apparent when inspecting the posterior distributions of the {SE 0.3, ET 0.1, LP $10^4$, MM off} run. 
Here we find a clear bi-modality: this inference run is able to correctly identify the more complex hot spot, however it also shows the presence of an additional secondary mode where the \texttt{PST} region is actually identified as an \texttt{ST} hot spot and vice-versa. 
This local maxima in the posterior surface seems to dominate the 
progression of the less computationally expensive run, which is therefore unable to reveal the additional, higher-likelihood mode. 

However, neither our runs is able to identify the mode associated with the correct solution. 
By checking the likelihood value correspondent to the injected parameter vector, we notice that in both cases the log-likelihood of the injected solution is greater than the maximum likelihood solution found by the sampler; however only by a factor of a few in $\log$, for the 
{SE 0.3, ET 0.1, LP $10^4$, MM off} run. 
This suggests that another mode is present in the posterior surface and that the relevant part of the parameter space has not yet been adequately explored and/or that the prior volume supporting the mode, found without external constraints, has now been significantly reduced, such that it is harder to identify it.  
Indeed neither inference run found samples space close to the injected vector.

Although our work here has revealed that systematics can occur in the \ac{PPM} analysis of \NICER data, it also highlights that these can be mitigated by convergence tests, proving that the parameter space has been exhaustively explored. 
These should include increasingly more computationally expensive runs, with more and more stringent sampling requirements, as well as repeated inferences, assessing the variability due to the random processes, and posterior predictive distribution tests.
Since we find that in general mass and radius are well recovered if no further constraints are added (even when the geometry parameters are not), our findings also suggest it may be beneficial to accompany inference runs with tight constraints on geometry parameters, when these are available, with runs that do not consider them. 
We are now prioritizing computer resources to ensure that we can carry out such targeted and comprehensive convergence tests on upcoming real data sets.

In the two \texttt{ST+PST} inference runs (with tight constraints on the inclination prior) here considered, 
applying background constraints would have not improved our findings, since the recovered background is always very similar to the injected one (see Figure \ref{fig:BKG}). 
This, however, is not necessarily the case for the real data. On the contrary, if the background constraints could cut the level of background found with these analyses, it would eliminate a large group of possible - and possibly similarly good - solutions, maybe uncovering a prominent but less ambiguous portion of the parameter space. 
\NICER background constraints have been applied on \NICER data sets for \joh \citep[][]{Salmi2022} and are currently being adopted for \NICER analyses on new, and expanded \NICER data sets for multiple \NICER sources. 

Similar constraints could also be provided through observations of \NICER sources by other X-ray (and in particular, imaging) telescopes. 
For example in \citet[][]{Riley2021, Miller2021}, the portion of the \NICER data attributed to the thermal emission of \joh, was constrained by the \xmm observations. 
Since coherently including \xmm data into \XPSI inference has been proven very beneficial, this procedure is also planned for other \NICER sources.

\subsection{Implications for PSR~J0030+0451}
Our study demonstrates that the analysis (\XPSI and \MultiNest) settings need to be tailored to each specific data set and applied assumptions. 
In particular our work places some of the findings, reported in \citetalias{Riley2019}, in a broader context. 
The new uncertainties and complications of the analysis process revealed in this study imply that \joo \NICER results and their interpretation need further investigations.  
Such studies are crucial to validate the robustness of the implications on the equation of state and the magnetic field structures derived from previous \joo \NICER works. 
This is the main target of the upcoming reanalysis of \NICER data \citep[][submitted]{Vinciguerra2023}.

\subsection{Implications for PSR~J0740+6620}
\joh is the second \NICER source, whose \ac{PPM} analyses have been publicly released by the \NICER collaboration. 
The inference of \joh's radius has crucial implications for the equation of state, given the very high mass, independently inferred from radio observations 
of $2.08 \pm0.07\,M_\odot$ 
\citep[][]{Fonseca20}. The same study has also provided meaningful constraints on distance and inclination. This information has been used in the \XPSI analyses of \joh \NICER data sets \citep[][]{Riley2021, Salmi2022}. 
While no simulation has yet been published to test the recovery performance of \XPSI for similar parameter vectors, we expect these studies to have delivered accurate results (T. Salmi et al. 2023, in preparation).  In \citet{Riley2021}, \joh was analysed with very different numbers of \MultiNest live points, proving stability in the solution found and the absence of other high-likelihood modes (a high number of live points, $4\times10^4$, was in the end used for production runs to correctly render the width of the posteriors; however no significant shift was found compared to runs with fewer live points). 
This more detailed analysis was possible for \joh, and not for \joo, because of its fewer counts, lower signal to noise ratio and the simpler model (the \texttt{ST-U} model was indeed identified as the headline model).  
 \subsection{Caveats} 
 \label{subsec:caveats}
 The simulations presented in this paper are far more exhaustive, and hence computationally expensive, than those done previously, but are still finite in scope. 
 From them we have learned that the noise greatly impacts our results, changing, in particular, the width of our posteriors as well as their sensitivity to the analysis settings. 
The extent of this effect however could not be fully inferred with our limited resources. 

 Our findings could also be significantly affected by the particular choices of parameter vectors and background spectra adopted to simulate the considered data sets.  
 Our sensitivity to such choices has not been tested here; 
 however the difference in behaviour found in this paper, compared to the results reported in \citetalias[][]{Riley2019} (e.g., in the effect of using different models on radius inference and evidence), suggests their impact could be significant. 
 This difference in behaviour could also lie only in the parameter vector used for the \texttt{ST+PST} model. 
 This parameter vector was indeed found in a low-resolution inference run \citep{Vinciguerra2023}; while data were then built with it at higher resolution. 
 This change might have produced some features in the simulated data sets which have no correspondence in the real data and may therefore explain the difference behaviour (in particular for the evidence comparison with different models) between simulated and real data \citepalias[see evidence discussion in Section 3 of ][]{Riley2019}.
Indeed when simulating the  \texttt{ST+PST} data, the inferred exposure time differs from the real one by $\sim 50\,$s, in contrast with only $0.01\,$s in the case of the \texttt{ST-U} model and parameter vector.

We also highlight here that in all the tests presented in this work, the physics used to produce the synthetic data sets was known and mostly (except for the \texttt{ST-U} inference runs on data produced with the \texttt{ST+PST} model) entirely captured by the inference set-up. 
This is not necessarily the case for the analysis of real data obtained by \NICER, where the physics is sometimes assumed (e.g., for the atmosphere composition) and sometimes approximated (e.g., for the specific hot spot shapes).

 \subsubsection{Correlations}
 To correctly interpret our results, we need to be aware of the various correlations between the parameter models. 
 For example, when we presented the percentage of parameters recovered within the $\sim 68\%$ credible interval and the relative uncertainties, 
 we had to assume that all the parameters were independent. 
 This is however not the case and can have considerable impact; for instance if a certain inclination is favoured by the sampling process, this will likely also shift the values of hot spot centers and sizes, as mentioned in Section \ref{subsec:degeneracies}. 
Moreover, as explained in the same section, we can obtain very similar emitting patterns with significantly different parameter values. 
For example, our tests seem to hint at weak sensitivity of our analysis to the smaller details of the hot spot shapes. 
Some emitting arcs could then be placed in either the two opposite directions without considerably changing the counts per channel and phase bin detected by \NICER. 
However the parameter values describing these two configurations would significantly differ from one another. 
These, at least partly, explain the poorer parameter recovery found for the tested \texttt{ST+PST} configuration.

Correlations should also be considered, when using the results presented by the \NICER collaboration. 
If one is interested in a single quantity, it is appropriate to use the median and credible interval reported for that 1D posterior of that model parameter (and marginalised over all the others).  
However, for reproducing a configuration that well represents the data, it is instead advisable to account for these correlations by selecting one (or more) appropriate specific sample(s). 
Tables \ref{tab:STU_diff_values} and \ref{tab:STPSTfromSTU_modes} show, with a specific example, how different the values of mass and radius can be for different modes and different samples, and how different they can be compared to properties describing their 1D posteriors, such as the mean. 
Even considering only the main mode, opting for the maximum likelihood sample or the maximum posterior one would make a difference to the \ac{NS} radius of almost 2\,km. 
These considerations are particularly relevant when the posteriors show multi-modal structures with similar probability and therefore figures describing the overall distributions as means and medians could take values that are totally inadequate (i.e. with very low posterior support) in reproducing the data. 

\section{Summary and Conclusions}\label{sec:conclusion}
This paper investigates the 
performance of \XPSI, one of the two main pipelines currently in use within the \NICER collaboration for \ac{PPM}.  
Simulation studies are particularly crucial to validate the results obtained for sources lacking external constraints, as it is the case for \joo. 
This study expands the work presented in \citet[][]{riley_thesis,Bogdanov2021}, by focusing on simulations that resemble \joo \NICER data with \texttt{ST-U} and, for the first time, \texttt{ST+PST} models. 
The former is the simplest model able to reproduce \joo \NICER data set with acceptable residuals  \citepalias[][]{Riley2019}; it describes each of the two hot spots with a spherical cap of uniform temperature. The latter was the model favored by the evidence in the study of \citetalias[][]{Riley2019}; compared to the \texttt{ST-U} model it introduces a third element which masks the emission from one of the two hot spots, giving it a more complex shape. 
This work presents the first investigation of parameter recovery for the \texttt{ST+PST} model, on which the headline results of 
\citetalias[][]{Riley2019} are based. 
We also study the impact of noise, analysis settings, external constraints, and lack/excess of model complexity. 
Below we list a summary of the most relevant lessons learnt: 
\begin{itemize}[noitemsep]
    \item Focusing on mass, radius, and compactness our findings validate the inference analyses performed by \XPSI for both models; 
    \item The overall parameter recovery performance of \XPSI for the \texttt{ST-U} inference runs is consistent with expectations and supports the results of \citet[][]{riley_thesis,Bogdanov2021}; 
    \item The overall parameter recovery from \texttt{ST+PST} runs is challenged by the increased complexity: degeneracies and correlations complicate the performance evaluation;
    \item For both models, the posterior surface is often characterised by a multi-modal structure. Possible future strategies to mitigate this challenge (once assured that the parameter space has been adequately explored) could include constraints based on independent findings (coming from additional \NICER sources, other phenomena or theoretical development) that could isolate the correct mode;
    \item The specific noise realisation can significantly impact the inference process. In particular it can considerably affect our sensitivity to settings and the widths of the posterior distributions. 
    There is therefore an additional source of scatter in the uncertainties on mass and radius inferences that can affect predictions such as the ones proposed by \citet[][]{Psaltis2014};
    \item As expected the noise realisation can also drive the best fitting solution away from the truth;
    \item We can potentially save computational resources by adopting the \XPSI low resolution settings (as described in Section \ref{sec:settings}) without compromising the  inference process;
    \item With the adopted settings and data sets, \MultiNest does not always adequately sample the parameter space to reveal all the maxima of the posterior surface (see \texttt{ST+PST} inference runs); residuals however do not show any prominent features. A sufficient exploration of the parameter space, through multiple runs with different analysis settings and based on simulated data, is therefore needed to assure the robustness of \XPSI results;
    \item In light of the uncovered multi-modal posterior surface often present in our inference analyses, evidences should be carefully evaluated. They also do not always help in identifying the most adequate model complexity (i.e. sometimes the difference in log-evidence between the \texttt{ST-U} and the \texttt{ST+PST} model is not significant);
    \item \joo-like data sets could be similarly reproduced by many diverse configurations, without showing any particular feature in the residuals; 
    \item There are a few discrepancies between the behaviour of the analyses performed on the simulated or real data sets. For example there is now a much smaller difference in log-evidence between the runs using \texttt{ST-U} and \texttt{ST+PST} models, and the mass and radius of a specific data set seem to be recovered independent of the model used for the inference analyses); both these findings differ from what was reported in \citetalias[][]{Riley2019}). Given an adequate amount of resources, a more comprehensive set of parameter vectors should be analysed; 
    \item As expected, introducing tight constraints on mass, distance and inclination can noticeably reduce the radius uncertainties;
    \item Because of the multi-modal structure of the posterior surface, applying tight constraints on parameter priors could potentially introduce biases in our results. In this work we see it clearly when we adopted mock constraints on \joo inclination. 
    The better likelihood associated with the injected parameter vector, however, suggests that more adequate sampling settings would allow the identification of the main mode, corresponding to the injected configuration;
    \item The tests done with increasing number of live points in \citet[][]{Riley2021} suggests that the radius inferences performed on \joh \NICER data sets are not affected by the same challenges identified here for the considered synthetic \joo-like data sets.
    
\end{itemize}
Our tests have therefore identified noise and multi-modal structure in the posterior (mostly due to degeneracies between the model parameters) as the two most prominent challenges of \ac{PPM} analyses conducted with \XPSI. 
Our findings also identified convergence tests, tailored to the specific data set and analysis of interest as a possible solution to both of them. 
These convergence tests will aim at assessing whether the parameter space is adequately explored and the uncovered posterior faithfully reflects the real one. They will include multiple runs with the same data set and model and increasingly stringent sampling settings (and in particular with an increasingly larger number of live points) and repeated runs to quantify the variability due to the randomness of the processes involved. We also plan to implement posterior predictive distribution checks and, on a longer time scale, to also adopt different and more sophisticated sampling algorithm \citep[such as UltraNest,][]{Buchner2021}. Although we will always be computationally limited, we think that these tests will help us to build a more solid interpretation of our results and get an overall understanding of the complexity of the posterior surface. 
Given the results presented in this work, we plan to accompany future analyses of \NICER data with a few inference runs on data simulated near the the recovered solution. These tests will require additional computational resources to ensure the robustness of \NICER findings on \ac{PPM}.

Despite the caveats listed in Section \ref{subsec:caveats}, this work shows that \XPSI recovers mass, radius and compactness according to expectations, when the settings guarantee the convergence of the sampling procedure.  

\begin{acknowledgments}
The authors thank Evert Rol and Martin Heemskerk for technical assistance. We acknowledge support from ERC Consolidator Grant No. 865768 AEONS (PI: Watts). This work was sponsored by NWO Domain Science for the use of the national computer facilities. Part of the work was carried out on the HELIOS cluster including dedicated nodes funded via the above mentioned ERC CoG.\\
{\it Facility}: \NICER \citep{Gendreau2016}.\\
{\it Software}: Python/C language \citep{Oliphant2007}, 
GNU Scientific Library (GSL; \citealt{Gough2009}), 
NumPy \citep{Walt2011}, 
Cython \citep{Behnel2011}, 
SciPy \citep{Jones},
OpenMP \citep{Dagum1998}, 
MPI \citep{Forum1994}, 
MPI for Python \citep{Dalcin2008},
Matplotlib \citep{Hunter2007, Droettboom2018}, 
IPython \citep{Perez2007}, 
Jupyter \citep{Kluyver2016}, 
TEMPO2 (photons; \citealt{Hobbs2006}), 
PINT (photonphase; https://github.com/ nanograv/PINT),
\MultiNest \citep{multinest09}, 
\PyMultiNest \citep{PyMultiNest}, GetDist (https://github.com/ cmbant/getdist), nestcheck \citep{Higson2018JOSS,Higson2018, Higson2019}, 
fgivenx \citep{Handley2018}, \XPSI (v1.0; https://github.com/ xpsi-group/xpsi; \citealt{xpsi}).

\end{acknowledgments}

\bibliographystyle{aasjournal}
\bibliography{allbib}

\begin{thebibliography}{}
\expandafter\ifx\csname natexlab\endcsname\relax\def\natexlab#1{#1}\fi
\providecommand{\url}[1]{\href{#1}{#1}}
\providecommand{\dodoi}[1]{doi:~\href{http://doi.org/#1}{\nolinkurl{#1}}}
\providecommand{\doeprint}[1]{\href{http://ascl.net/#1}{\nolinkurl{http://ascl.net/#1}}}
\providecommand{\doarXiv}[1]{\href{https://arxiv.org/abs/#1}{\nolinkurl{https://arxiv.org/abs/#1}}}

\bibitem[{{Afle} {et~al.}(2023){Afle}, {Miles}, {Caino-Lores}, {Capano},
  {Tews}, {Vahi}, {Deelman}, {Taufer}, \& {Brown}}]{Afle23}
{Afle}, C., {Miles}, P.~R., {Caino-Lores}, S., {et~al.} 2023, arXiv e-prints,
  arXiv:2304.01035, \dodoi{10.48550/arXiv.2304.01035}

\bibitem[{{Arons}(1981)}]{arons81}
{Arons}, J. 1981, \apj, 248, 1099, \dodoi{10.1086/159239}

\bibitem[{Ashton {et~al.}(2019)Ashton, H{\"u}bner, Lasky, Talbot, Ackley,
  Biscoveanu, Chu, Divakarla, Easter, Goncharov, {et~al.}}]{bilby2019}
Ashton, G., H{\"u}bner, M., Lasky, P.~D., {et~al.} 2019, The Astrophysical
  Journal Supplement Series, 241, 27

\bibitem[{{Baym} {et~al.}(2018){Baym}, {Hatsuda}, {Kojo}, {Powell}, {Song}, \&
  {Takatsuka}}]{Baym2018}
{Baym}, G., {Hatsuda}, T., {Kojo}, T., {et~al.} 2018, Reports on Progress in
  Physics, 81, 056902, \dodoi{10.1088/1361-6633/aaae14}

\bibitem[{{Behnel} {et~al.}(2011){Behnel}, {Bradshaw}, {Citro}, {Dalcin},
  {Seljebotn}, \& {Smith}}]{Behnel2011}
{Behnel}, S., {Bradshaw}, R., {Citro}, C., {et~al.} 2011, Computing in Science
  and Engineering, 13, 31, \dodoi{10.1109/MCSE.2010.118}

\bibitem[{{Berry} {et~al.}(2015){Berry}, {Mandel}, {Middleton}, {Singer},
  {Urban}, {Vecchio}, {Vitale}, {Cannon}, {Farr}, {Farr}, {Graff}, {Hanna},
  {Haster}, {Mohapatra}, {Pankow}, {Price}, {Sidery}, \& {Veitch}}]{Berry2015}
{Berry}, C. P.~L., {Mandel}, I., {Middleton}, H., {et~al.} 2015, \apj, 804,
  114, \dodoi{10.1088/0004-637X/804/2/114}

\bibitem[{{Bilous} {et~al.}(2019){Bilous}, {Watts}, {Harding}, {Riley},
  {Arzoumanian}, {Bogdanov}, {Gendreau}, {Ray}, {Guillot}, {Ho}, \&
  {Chakrabarty}}]{Bilous_2019}
{Bilous}, A.~V., {Watts}, A.~L., {Harding}, A.~K., {et~al.} 2019, \apjl, 887,
  L23, \dodoi{10.3847/2041-8213/ab53e7}

\bibitem[{{Bogdanov} {et~al.}(2019{\natexlab{a}}){Bogdanov}, {Guillot}, {Ray},
  {et~al.}}]{bogdanov19a}
{Bogdanov}, S., {Guillot}, S., {Ray}, P.~S., {et~al.} 2019{\natexlab{a}},
  \apjl, 887, L25, \dodoi{10.3847/2041-8213/ab53eb}

\bibitem[{{Bogdanov} {et~al.}(2019{\natexlab{b}}){Bogdanov}, {Lamb},
  {Mahmoodifar}, {Miller}, {Morsink}, {Riley}, {Strohmayer}, {Tung}, {Watts},
  {Dittmann}, {Chakrabarty}, {Guillot}, {Arzoumanian}, \&
  {Gendreau}}]{Bogdanov2019b}
{Bogdanov}, S., {Lamb}, F.~K., {Mahmoodifar}, S., {et~al.} 2019{\natexlab{b}},
  \apjl, 887, L26, \dodoi{10.3847/2041-8213/ab5968}

\bibitem[{{Bogdanov} {et~al.}(2021){Bogdanov}, {Dittmann}, {Ho}, {Lamb},
  {Mahmoodifar}, {Miller}, {Morsink}, {Riley}, {Strohmayer}, {Watts},
  {Choudhury}, {Guillot}, {Harding}, {Ray}, {Wadiasingh}, {Wolff}, {Markwardt},
  {Arzoumanian}, \& {Gendreau}}]{Bogdanov2021}
{Bogdanov}, S., {Dittmann}, A.~J., {Ho}, W. C.~G., {et~al.} 2021, \apjl, 914,
  L15, \dodoi{10.3847/2041-8213/abfb79}

\bibitem[{{Buchner}(2021)}]{Buchner2021}
{Buchner}, J. 2021, The Journal of Open Source Software, 6, 3001,
  \dodoi{10.21105/joss.03001}

\bibitem[{{Buchner} {et~al.}(2014{\natexlab{a}}){Buchner}, {Georgakakis},
  {Nandra}, {Hsu}, {Rangel}, {Brightman}, {Merloni}, {Salvato}, {Donley}, \&
  {Kocevski}}]{PyMultiNest}
{Buchner}, J., {Georgakakis}, A., {Nandra}, K., {et~al.} 2014{\natexlab{a}},
  \aap, 564, A125, \dodoi{10.1051/0004-6361/201322971}

\bibitem[{{Buchner} {et~al.}(2014{\natexlab{b}}){Buchner}, {Georgakakis},
  {Nandra}, {Hsu}, {Rangel}, {Brightman}, {Merloni}, {Salvato}, {Donley}, \&
  {Kocevski}}]{Buchner2014}
---. 2014{\natexlab{b}}, \aap, 564, A125, \dodoi{10.1051/0004-6361/201322971}

\bibitem[{{Cameron}(2011)}]{Cameron:2011}
{Cameron}, E. 2011, \pasa, 28, 128, \dodoi{10.1071/AS10046}

\bibitem[{{Chen} {et~al.}(2020){Chen}, {Yuan}, \& {Vasilopoulos}}]{Chen2020}
{Chen}, A.~Y., {Yuan}, Y., \& {Vasilopoulos}, G. 2020, \apjl, 893, L38,
  \dodoi{10.3847/2041-8213/ab85c5}

\bibitem[{{Cromartie} {et~al.}(2020){Cromartie}, {Fonseca}, {Ransom},
  {Demorest}, {Arzoumanian}, {Blumer}, {Brook}, {DeCesar}, {Dolch}, {Ellis},
  {Ferdman}, {Ferrara}, {Garver-Daniels}, {Gentile}, {Jones}, {Lam}, {Lorimer},
  {Lynch}, {McLaughlin}, {Ng}, {Nice}, {Pennucci}, {Spiewak}, {Stairs},
  {Stovall}, {Swiggum}, \& {Zhu}}]{Cromartie2020}
{Cromartie}, H.~T., {Fonseca}, E., {Ransom}, S.~M., {et~al.} 2020, Nature
  Astronomy, 4, 72, \dodoi{10.1038/s41550-019-0880-2}

\bibitem[{Dagum \& Menon(1998)}]{Dagum1998}
Dagum, L., \& Menon, R. 1998, IEEE Computational Science and Engineering, 5,
  46, \dodoi{10.1109/99.660313}

\bibitem[{Dalcín {et~al.}(2008)Dalcín, Paz, Storti, \&
  D’Elía}]{Dalcin2008}
Dalcín, L., Paz, R., Storti, M., \& D’Elía, J. 2008, Journal of Parallel
  and Distributed Computing, 68, 655,
  \dodoi{https://doi.org/10.1016/j.jpdc.2007.09.005}

\bibitem[{{Feroz} \& {Hobson}(2008)}]{Feroz2008}
{Feroz}, F., \& {Hobson}, M.~P. 2008, \mnras, 384, 449,
  \dodoi{10.1111/j.1365-2966.2007.12353.x}

\bibitem[{{Feroz} {et~al.}(2009{\natexlab{a}}){Feroz}, {Hobson}, \&
  {Bridges}}]{Feroz2009}
{Feroz}, F., {Hobson}, M.~P., \& {Bridges}, M. 2009{\natexlab{a}}, \mnras, 398,
  1601, \dodoi{10.1111/j.1365-2966.2009.14548.x}

\bibitem[{{Feroz} {et~al.}(2009{\natexlab{b}}){Feroz}, {Hobson}, \&
  {Bridges}}]{multinest09}
---. 2009{\natexlab{b}}, \mnras, 398, 1601,
  \dodoi{10.1111/j.1365-2966.2009.14548.x}

\bibitem[{{Feroz} {et~al.}(2019){Feroz}, {Hobson}, {Cameron}, \&
  {Pettitt}}]{Feroz2019}
{Feroz}, F., {Hobson}, M.~P., {Cameron}, E., \& {Pettitt}, A.~N. 2019, The Open
  Journal of Astrophysics, 2, 10, \dodoi{10.21105/astro.1306.2144}

\bibitem[{{Fonseca} {et~al.}(2021){Fonseca}, {Cromartie}, {Pennucci}, {Ray},
  {Kirichenko}, {Ransom}, {Demorest}, {Stairs}, {Arzoumanian}, {Guillemot},
  {Parthasarathy}, {Kerr}, {Cognard}, {Baker}, {Blumer}, {Brook}, {DeCesar},
  {Dolch}, {Dong}, {Ferrara}, {Fiore}, {Garver-Daniels}, {Good}, {Jennings},
  {Jones}, {Kaspi}, {Lam}, {Lorimer}, {Luo}, {McEwen}, {McKee}, {McLaughlin},
  {McMann}, {Meyers}, {Naidu}, {Ng}, {Nice}, {Pol}, {Radovan},
  {Shapiro-Albert}, {Tan}, {Tendulkar}, {Swiggum}, {Wahl}, \&
  {Zhu}}]{Fonseca20}
{Fonseca}, E., {Cromartie}, H.~T., {Pennucci}, T.~T., {et~al.} 2021, \apjl,
  915, L12, \dodoi{10.3847/2041-8213/ac03b8}

\bibitem[{Forum(1994)}]{Forum1994}
Forum, M.~P. 1994, MPI: A Message-Passing Interface Standard,
  \url{http://www.scipy.org/}

\bibitem[{{Gendreau} {et~al.}(2016){Gendreau}, {Arzoumanian}, {Adkins},
  {Albert}, {Anders}, {Aylward}, {Baker}, {Balsamo}, {Bamford}, {Benegalrao},
  {Berry}, {Bhalwani}, {Black}, {Blaurock}, {Bronke}, {Brown}, {Budinoff},
  {Cantwell}, {Cazeau}, {Chen}, {Clement}, {Colangelo}, {Coleman},
  {Coopersmith}, {Dehaven}, {Doty}, {Egan}, {Enoto}, {Fan}, {Ferro}, {Foster},
  {Galassi}, {Gallo}, {Green}, {Grosh}, {Ha}, {Hasouneh}, {Heefner}, {Hestnes},
  {Hoge}, {Jacobs}, {J{\o}rgensen}, {Kaiser}, {Kellogg}, {Kenyon}, {Koenecke},
  {Kozon}, {LaMarr}, {Lambertson}, {Larson}, {Lentine}, {Lewis}, {Lilly},
  {Liu}, {Malonis}, {Manthripragada}, {Markwardt}, {Matonak}, {Mcginnis},
  {Miller}, {Mitchell}, {Mitchell}, {Mohammed}, {Monroe}, {Montt de Garcia},
  {Mul{\'e}}, {Nagao}, {Ngo}, {Norris}, {Norwood}, {Novotka}, {Okajima},
  {Olsen}, {Onyeachu}, {Orosco}, {Peterson}, {Pevear}, {Pham}, {Pollard},
  {Pope}, {Powers}, {Powers}, {Price}, {Prigozhin}, {Ramirez}, {Reid},
  {Remillard}, {Rogstad}, {Rosecrans}, {Rowe}, {Sager}, {Sanders}, {Savadkin},
  {Saylor}, {Schaeffer}, {Schweiss}, {Semper}, {Serlemitsos}, {Shackelford},
  {Soong}, {Struebel}, {Vezie}, {Villasenor}, {Winternitz}, {Wofford},
  {Wright}, {Yang}, \& {Yu}}]{Gendreau2016}
{Gendreau}, K.~C., {Arzoumanian}, Z., {Adkins}, P.~W., {et~al.} 2016, in
  Society of Photo-Optical Instrumentation Engineers (SPIE) Conference Series,
  Vol. 9905, Space Telescopes and Instrumentation 2016: Ultraviolet to Gamma
  Ray, ed. J.-W.~A. {den Herder}, T.~{Takahashi}, \& M.~{Bautz}, 99051H,
  \dodoi{10.1117/12.2231304}

\bibitem[{{Gough}(2009)}]{Gough2009}
{Gough}, B. 2009, {GNU Scientific Library Reference Manual} (Network Theory
  Ltd.)

\bibitem[{{Handley}(2018)}]{Handley2018}
{Handley}, W. 2018, The Journal of Open Source Software, 3, 849,
  \dodoi{10.21105/joss.00849}

\bibitem[{{Harding} \& {Muslimov}(2001)}]{HM01}
{Harding}, A.~K., \& {Muslimov}, A.~G. 2001, \apj, 556, 987,
  \dodoi{10.1086/321589}

\bibitem[{{Hebeler}(2021)}]{Hebeler2021}
{Hebeler}, K. 2021, \physrep, 890, 1, \dodoi{10.1016/j.physrep.2020.08.009}

\bibitem[{{Higson}(2018)}]{Higson2018JOSS}
{Higson}, E. 2018, The Journal of Open Source Software, 3, 916,
  \dodoi{10.21105/joss.00916}

\bibitem[{{Higson} {et~al.}(2018){Higson}, {Handley}, {Hobson}, \&
  {Lasenby}}]{Higson2018}
{Higson}, E., {Handley}, W., {Hobson}, M., \& {Lasenby}, A. 2018, Bayesian
  Analysis, 13, 873, \dodoi{10.1214/17-BA1075}

\bibitem[{{Higson} {et~al.}(2019){Higson}, {Handley}, {Hobson}, \&
  {Lasenby}}]{Higson2019}
---. 2019, \mnras, 483, 2044, \dodoi{10.1093/mnras/sty3090}

\bibitem[{{Ho} \& {Heinke}(2009)}]{HH09}
{Ho}, W.~C.~G., \& {Heinke}, C.~O. 2009, \nat, 462, 71,
  \dodoi{10.1038/nature08525}

\bibitem[{{Ho} \& {Lai}(2001)}]{Ho01}
{Ho}, W. C.~G., \& {Lai}, D. 2001, \mnras, 327, 1081,
  \dodoi{10.1046/j.1365-8711.2001.04801.x}

\bibitem[{{Hobbs} {et~al.}(2006){Hobbs}, {Edwards}, \&
  {Manchester}}]{Hobbs2006}
{Hobbs}, G.~B., {Edwards}, R.~T., \& {Manchester}, R.~N. 2006, \mnras, 369,
  655, \dodoi{10.1111/j.1365-2966.2006.10302.x}

\bibitem[{Hunter(2007)}]{Hunter2007}
Hunter, J.~D. 2007, Computing in Science \& Engineering, 9, 90,
  \dodoi{10.1109/MCSE.2007.55}

\bibitem[{Jones {et~al.}(2001)Jones, Oliphant, Peterson, \& et~al.}]{Jones}
Jones, E., Oliphant, T., Peterson, P., \& et~al. 2001, SciPy: Open source
  scientific tools for Python., \url{http://www.scipy.org/}

\bibitem[{{Kalapotharakos} {et~al.}(2021){Kalapotharakos}, {Wadiasingh},
  {Harding}, \& {Kazanas}}]{Kalapotharakos2021}
{Kalapotharakos}, C., {Wadiasingh}, Z., {Harding}, A.~K., \& {Kazanas}, D.
  2021, \apj, 907, 63, \dodoi{10.3847/1538-4357/abcec0}

\bibitem[{{Kluyver} {et~al.}(2016){Kluyver}, {Ragan-Kelley}, {P{\'e}rez},
  {Granger}, {Bussonnier}, {Frederic}, {Kelley}, {Hamrick}, {Grout}, {Corlay},
  {Ivanov}, {Avila}, {Abdalla}, {Willing}, \& {Jupyter Development
  Team}}]{Kluyver2016}
{Kluyver}, T., {Ragan-Kelley}, B., {P{\'e}rez}, F., {et~al.} 2016, in IOS
  Press, 87--90, \dodoi{10.3233/978-1-61499-649-1-87}

\bibitem[{{Lattimer}(2012)}]{Lattimer12ARNPS}
{Lattimer}, J.~M. 2012, Annual Review of Nuclear and Particle Science, 62, 485,
  \dodoi{10.1146/annurev-nucl-102711-095018}

\bibitem[{{Lo} {et~al.}(2013){Lo}, {Miller}, {Bhattacharyya}, \&
  {Lamb}}]{lomiller13}
{Lo}, K.~H., {Miller}, M.~C., {Bhattacharyya}, S., \& {Lamb}, F.~K. 2013, \apj,
  776, 19, \dodoi{10.1088/0004-637X/776/1/19}

\bibitem[{Michael {et~al.}(2018)Michael, Caswell, Hunter, Firing, \&
  et~al.}]{Droettboom2018}
Michael, D., Caswell, T.~A., Hunter, J., Firing, E., \& et~al. 2018,
  {matplotlib/matplotlib v2.2.2}, v2.2.2,  Zenodo,
  \dodoi{10.5281/zenodo.1202077}

\bibitem[{{Miller} \& {Lamb}(2015)}]{Miller2015}
{Miller}, M.~C., \& {Lamb}, F.~K. 2015, \apj, 808, 31,
  \dodoi{10.1088/0004-637X/808/1/31}

\bibitem[{{Miller} {et~al.}(2019){Miller}, {Lamb}, {Dittmann}, {Bogdanov},
  {Arzoumanian}, {Gendreau}, {Guillot}, {Harding}, {Ho}, {Lattimer}, {Ludlam},
  {Mahmoodifar}, {Morsink}, {Ray}, {Strohmayer}, {Wood}, {Enoto}, {Foster},
  {Okajima}, {Prigozhin}, \& {Soong}}]{Miller2019}
{Miller}, M.~C., {Lamb}, F.~K., {Dittmann}, A.~J., {et~al.} 2019, \apjl, 887,
  L24, \dodoi{10.3847/2041-8213/ab50c5}

\bibitem[{{Miller} {et~al.}(2021){Miller}, {Lamb}, {Dittmann}, {Bogdanov},
  {Arzoumanian}, {Gendreau}, {Guillot}, {Ho}, {Lattimer}, {Loewenstein},
  {Morsink}, {Ray}, {Wolff}, {Baker}, {Cazeau}, {Manthripragada}, {Markwardt},
  {Okajima}, {Pollard}, {Cognard}, {Cromartie}, {Fonseca}, {Guillemot}, {Kerr},
  {Parthasarathy}, {Pennucci}, {Ransom}, \& {Stairs}}]{Miller2021}
---. 2021, \apjl, 918, L28, \dodoi{10.3847/2041-8213/ac089b}

\bibitem[{{Morsink} {et~al.}(2007){Morsink}, {Leahy}, {Cadeau}, \&
  {Braga}}]{Morsink2007}
{Morsink}, S.~M., {Leahy}, D.~A., {Cadeau}, C., \& {Braga}, J. 2007, \apj, 663,
  1244, \dodoi{10.1086/518648}

\bibitem[{{Oertel} {et~al.}(2017){Oertel}, {Hempel}, {Kl{\"a}hn}, \&
  {Typel}}]{OE17}
{Oertel}, M., {Hempel}, M., {Kl{\"a}hn}, T., \& {Typel}, S. 2017, Reviews of
  Modern Physics, 89, 015007, \dodoi{10.1103/RevModPhys.89.015007}

\bibitem[{{Oliphant}(2007)}]{Oliphant2007}
{Oliphant}, T.~E. 2007, Computing in Science and Engineering, 9, 10,
  \dodoi{10.1109/MCSE.2007.58}

\bibitem[{Perez \& Granger(2007)}]{Perez2007}
Perez, F., \& Granger, B.~E. 2007, Computing in Science \& Engineering, 9, 21,
  \dodoi{10.1109/MCSE.2007.53}

\bibitem[{{Psaltis} {et~al.}(2014){Psaltis}, {{\"O}zel}, \&
  {Chakrabarty}}]{Psaltis2014}
{Psaltis}, D., {{\"O}zel}, F., \& {Chakrabarty}, D. 2014, \apj, 787, 136,
  \dodoi{10.1088/0004-637X/787/2/136}

\bibitem[{{Raaijmakers} {et~al.}(2021){Raaijmakers}, {Greif}, {Hebeler},
  {Hinderer}, {Nissanke}, {Schwenk}, {Riley}, {Watts}, {Lattimer}, \&
  {Ho}}]{Raaijmakers2021}
{Raaijmakers}, G., {Greif}, S.~K., {Hebeler}, K., {et~al.} 2021, \apjl, 918,
  L29, \dodoi{10.3847/2041-8213/ac089a}

\bibitem[{{Ray} {et~al.}(2019){Ray}, {Arzoumanian}, {Ballantyne}, {Bozzo},
  {Brandt}, {Brenneman}, {Chakrabarty}, {Christophersen}, {DeRosa}, {Feroci},
  {Gendreau}, {Goldstein}, {Hartmann}, {Hernanz}, {Jenke}, {Kara}, {Maccarone},
  {McDonald}, {Nowak}, {Phlips}, {Remillard}, {Stevens}, {Tomsick}, {Watts},
  {Wilson-Hodge}, {Wood}, {Zane}, {Ajello}, {Alston}, {Altamirano}, {Antoniou},
  {Arur}, {Ashton}, {Auchettl}, {Ayres}, {Bachetti}, {Balokovic}, {Baring},
  {Baykal}, {Begelman}, {Bhat}, {Bogdanov}, {Briggs}, {Bulbul}, {Bult},
  {Burns}, {Cackett}, {Campana}, {Caspi}, {Cavecchi}, {Chenevez}, {Cherry},
  {Corbet}, {Corcoran}, {Corsi}, {Degenaar}, {Drake}, {Eikenberry}, {Enoto},
  {Fragile}, {Fuerst}, {Gandhi}, {Garcia}, {Goldstein}, {Gonzalez},
  {Grefenstette}, {Grinberg}, {Grossan}, {Guillot}, {Guver}, {Haggard},
  {Heinke}, {Heinz}, {Hemphill}, {Homan}, {Hui}, {Huppenkothen}, {Ingram},
  {Irwin}, {Jaisawal}, {Jaodand}, {Kalemci}, {Kaplan}, {Keek}, {Kennea},
  {Kerr}, {van der Klis}, {Kocevski}, {Koss}, {Kowalski}, {Lai}, {Lamb},
  {Laycock}, {Lazio}, {Lazzati}, {Longcope}, {Loewenstein}, {Maitra}, {Majid},
  {Maksym}, {Malacaria}, {Margutti}, {Martindale}, {McHardy}, {Meyer},
  {Middleton}, {Miller}, {Miller}, {Motta}, {Neilsen}, {Nelson}, {Noble},
  {O'Brien}, {Osborne}, {Osten}, {Ozel}, {Palliyaguru}, {Pasham}, {Patruno},
  {Pelassa}, {Petropoulou}, {Pilia}, {Pohl}, {Pooley}, {Prescod-Weinstein},
  {Psaltis}, {Raaijmakers}, {Reynolds}, {Riley}, {Salvesen}, {Santangelo},
  {Scaringi}, {Schanne}, {Schnittman}, {Smith}, {Smith}, {Snios}, {Steiner},
  {Steiner}, {Stella}, {Strohmayer}, {Sun}, {Tauris}, {Taylor}, {Tohuvavohu},
  {Vacchi}, {Vasilopoulos}, {Veledina}, {Walsh}, {Weinberg}, {Wilkins},
  {Willingale}, {Wilms}, {Winter}, {Wolff}, {in 't Zand}, {Zezas}, {Zhang}, \&
  {Zoghbi}}]{strobex}
{Ray}, P.~S., {Arzoumanian}, Z., {Ballantyne}, D., {et~al.} 2019, arXiv
  e-prints, arXiv:1903.03035.
\newblock \doarXiv{1903.03035}

\bibitem[{{Riley}(2019)}]{riley_thesis}
{Riley}, T.~E. 2019, PhD thesis, University of Amsterdam.
\newblock
  \url{https://hdl.handle.net/11245.1/aa86fcf3-2437-4bc2-810e-cf9f30a98f7a}

\bibitem[{{Riley} {et~al.}(2018){Riley}, {Raaijmakers}, \& {Watts}}]{Riley2018}
{Riley}, T.~E., {Raaijmakers}, G., \& {Watts}, A.~L. 2018, \mnras, 478, 1093,
  \dodoi{10.1093/mnras/sty1051}

\bibitem[{{Riley} {et~al.}(2019){Riley}, {Watts}, {Bogdanov}, {Ray}, {Ludlam},
  {Guillot}, {Arzoumanian}, {Baker}, {Bilous}, {Chakrabarty}, {Gendreau},
  {Harding}, {Ho}, {Lattimer}, {Morsink}, \& {Strohmayer}}]{Riley2019}
{Riley}, T.~E., {Watts}, A.~L., {Bogdanov}, S., {et~al.} 2019, \apjl, 887, L21,
  \dodoi{10.3847/2041-8213/ab481c}

\bibitem[{{Riley} {et~al.}(2021){Riley}, {Watts}, {Ray}, {Bogdanov}, {Guillot},
  {Morsink}, {Bilous}, {Arzoumanian}, {Choudhury}, {Deneva}, {Gendreau},
  {Harding}, {Ho}, {Lattimer}, {Loewenstein}, {Ludlam}, {Markwardt}, {Okajima},
  {Prescod-Weinstein}, {Remillard}, {Wolff}, {Fonseca}, {Cromartie}, {Kerr},
  {Pennucci}, {Parthasarathy}, {Ransom}, {Stairs}, {Guillemot}, \&
  {Cognard}}]{Riley2021}
{Riley}, T.~E., {Watts}, A.~L., {Ray}, P.~S., {et~al.} 2021, \apjl, 918, L27,
  \dodoi{10.3847/2041-8213/ac0a81}

\bibitem[{{Riley} {et~al.}(2023){Riley}, {Choudhury}, {Salmi}, {Vinciguerra},
  {Kini}, {Dorsman}, {Watts}, {Huppenkothen}, \& {Guillot}}]{xpsi}
{Riley}, T.~E., {Choudhury}, D., {Salmi}, T., {et~al.} 2023, {Journal of Open
  Source Software}, 8, 4977, \dodoi{10.21105/joss.04977}

\bibitem[{{Ruderman} \& {Sutherland}(1975)}]{RudermanSutherland1975}
{Ruderman}, M.~A., \& {Sutherland}, P.~G. 1975, \apj, 196, 51,
  \dodoi{10.1086/153393}

\bibitem[{{Salmi} {et~al.}(2022){Salmi}, {Vinciguerra}, {Choudhury}, {Riley},
  {Watts}, {Remillard}, {Ray}, {Bogdanov}, {Guillot}, {Arzoumanian},
  {Chirenti}, {Dittmann}, {Gendreau}, {Ho}, {Miller}, {Morsink}, {Wadiasingh},
  \& {Wolff}}]{Salmi2022}
{Salmi}, T., {Vinciguerra}, S., {Choudhury}, D., {et~al.} 2022, \apj, 941, 150,
  \dodoi{10.3847/1538-4357/ac983d}

\bibitem[{{Skilling}(2004)}]{Skilling2004}
{Skilling}, J. 2004, in American Institute of Physics Conference Series, Vol.
  735, Bayesian Inference and Maximum Entropy Methods in Science and
  Engineering: 24th International Workshop on Bayesian Inference and Maximum
  Entropy Methods in Science and Engineering, ed. R.~{Fischer}, R.~{Preuss}, \&
  U.~V. {Toussaint}, 395--405, \dodoi{10.1063/1.1835238}

\bibitem[{{Tolos} \& {Fabbietti}(2020)}]{Tolos2020}
{Tolos}, L., \& {Fabbietti}, L. 2020, Progress in Particle and Nuclear Physics,
  112, 103770, \dodoi{10.1016/j.ppnp.2020.103770}

\bibitem[{{van der Walt} {et~al.}(2011){van der Walt}, {Colbert}, \&
  {Varoquaux}}]{Walt2011}
{van der Walt}, S., {Colbert}, S.~C., \& {Varoquaux}, G. 2011, Computing in
  Science and Engineering, 13, 22, \dodoi{10.1109/MCSE.2011.37}

\bibitem[{Vinciguerra {et~al.}(2023)Vinciguerra, Salmi, Watts, Choudhury, Kini,
  \& Riley}]{SIMUJ0030zenodo}
Vinciguerra, S., Salmi, T., Watts, A.~L., {et~al.} 2023, {X-PSI Parameter
  Recovery for Temperature Map Configurations Inspired by PSR J0030+0451},
  Zenodo, \dodoi{10.5281/zenodo.7646352}

\bibitem[{{Vinciguerra} {et~al.}(2023){Vinciguerra}, {Salmi}, {Watts},
  {Choudhury}, {Riley}, {Ray}, {Bogdanov}, {Kini}, {Guillot}, {Chakrabarty},
  {Ho}, {Huppenkothen}, {Morsink}, {Wadiasingh}, \& {Wolff}}]{Vinciguerra2023}
{Vinciguerra}, S., {Salmi}, T., {Watts}, L.~A., {et~al.} 2023, \apj, submitted

\bibitem[{{Watts}(2019)}]{Watts2019}
{Watts}, A.~L. 2019, in American Institute of Physics Conference Series, Vol.
  2127, Xiamen-CUSTIPEN Workshop on the Equation of State of Dense Neutron-Rich
  Matter in the Era of Gravitational Wave Astronomy, 020008,
  \dodoi{10.1063/1.5117798}

\bibitem[{{Watts} {et~al.}(2016){Watts}, {Andersson}, {Chakrabarty}, {Feroci},
  {Hebeler}, {Israel}, {Lamb}, {Miller}, {Morsink}, {{\"O}zel}, {Patruno},
  {Poutanen}, {Psaltis}, {Schwenk}, {Steiner}, {Stella}, {Tolos}, \& {van der
  Klis}}]{2016RvMP...88b1001W}
{Watts}, A.~L., {Andersson}, N., {Chakrabarty}, D., {et~al.} 2016, Reviews of
  Modern Physics, 88, 021001, \dodoi{10.1103/RevModPhys.88.021001}

\bibitem[{{Watts} {et~al.}(2019){Watts}, {Yu}, {Poutanen}, {Zhang},
  {Bhattacharyya}, {Bogdanov}, {Ji}, {Patruno}, {Riley}, {Bakala}, {Baykal},
  {Bernardini}, {Bombaci}, {Brown}, {Cavecchi}, {Chakrabarty}, {Chenevez},
  {Degenaar}, {Del Santo}, {Di Salvo}, {Doroshenko}, {Falanga}, {Ferdman},
  {Feroci}, {Gambino}, {Ge}, {Greif}, {Guillot}, {Gungor}, {Hartmann},
  {Hebeler}, {Heger}, {Homan}, {Iaria}, {Zand}, {Kargaltsev}, {Kurkela}, {Lai},
  {Li}, {Li}, {Li}, {Linares}, {Lu}, {Mahmoodifar}, {M{\'e}ndez}, {Coleman
  Miller}, {Morsink}, {N{\"a}ttil{\"a}}, {Possenti}, {Prescod-Weinstein}, {Qu},
  {Riggio}, {Salmi}, {Sanna}, {Santangelo}, {Schatz}, {Schwenk}, {Song}, {{\v
  S}r{\'a}mkov{\'a}}, {Stappers}, {Stiele}, {Strohmayer}, {Tews}, {Tolos},
  {T{\"o}r{\"o}k}, {Tsang}, {Urbanec}, {Vacchi}, {Xu}, {Xu}, {Zane}, {Zhang},
  {Zhang}, {Zhang}, {Zheng}, \& {Zhou}}]{dmatter_extp}
{Watts}, A.~L., {Yu}, W., {Poutanen}, J., {et~al.} 2019, Science China Physics,
  Mechanics, and Astronomy, 62, 29503, \dodoi{10.1007/s11433-017-9188-4}

\bibitem[{{Wilms} {et~al.}(2000){Wilms}, {Allen}, \& {McCray}}]{Wilms2000}
{Wilms}, J., {Allen}, A., \& {McCray}, R. 2000, \apj, 542, 914,
  \dodoi{10.1086/317016}

\bibitem[{{Yang} \& {Piekarewicz}(2020)}]{YangPiek2020}
{Yang}, J., \& {Piekarewicz}, J. 2020, Annual Review of Nuclear and Particle
  Science, 70, 21, \dodoi{10.1146/annurev-nucl-101918-023608}

\end{thebibliography}

\end{document}